\documentclass[onecolumn,smallcondensed,smallextended,dvips]{svjour3}
\usepackage{mathptmx}
\usepackage{color}

\usepackage{epsfig,graphics}
\long\def\beginpgfgraphicnamed#1#2\endpgfgraphicnamed{\includegraphics{#1}}

\usepackage{microtype}
\usepackage{fixltx2e}
\usepackage{calc}

\usepackage[centertags,intlimits]{amsmath}
\usepackage{amssymb,amsfonts}
\usepackage{mathtools}

\DeclareMathOperator{\e}{e}

\newcommand{\gbeta}{g^{(\beta)}}

\newcommand{\proba}[2]{\mathbb{P}_{#1}\left(#2\right)}

\journalname{Journal of Statistical Physics}

\begin{document}

\title{\large\bfseries The speed of evolution in large asexual populations}
\author{Su-Chan Park \and Damien Simon \and Joachim Krug}
\institute{S.-C. Park \and J. Krug \at
           Institute for Theoretical Physics, University of Cologne, Germany \\
           \email{psc@thp.uni-koeln.de, krug@thp.uni-koeln.de} \\
D. Simon \at
Universit\'e Pierre et Marie Curie,
Laboratoire ``Probabilit\'es et mod\`eles al\'eatoires'', Paris,
France \\ \email{damien.simon@normalesup.org}}          
\date{Received: date / Accepted: date}
\maketitle

\begin{abstract}
We consider an asexual biological population of constant size $N$ evolving 
in discrete time under the influence of selection and mutation. 
Beneficial mutations appear at rate $U$ and
their selective effects $s$ are drawn from a distribution $g(s)$.
After introducing the required models and concepts of mathematical
population genetics, we review different approaches to computing the 
speed of logarithmic fitness increase  as a function of $N$, $U$ and 
$g(s)$. We present an exact solution of the infinite population size limit and
provide an estimate of the population size beyond which it is valid. 
We then discuss approximate approaches to the finite population problem, distinguishing
between the case of a single selection coefficient, $g(s) = \delta(s - s_b)$, and a continuous
distribution of selection coefficients. Analytic estimates for the speed are compared to 
numerical simulations up to population sizes of order $10^{300}$.  
  \keywords{evolutionary dynamics; Wright-Fisher model; clonal interference; traveling waves}
\end{abstract}

\section{Introduction}

The foundations of mathematical population genetics were established
around 1930 in three seminal works of R. A. Fisher \cite{F1930},
J.B.S. Haldane \cite{Haldane1932} and S. Wright \cite{W1931}. The
achievement of these three pioneers is often referred to as the 
\textit{modern synthesis}, because they
resolved an apparent contradiction between Darwinian evolutionary
theory, with its emphasis on minute changes accumulating over long
times, and the then recently rediscovered laws of Mendelian genetics, 
which showed that
the hereditary material underlying these changes is intrinsically
discrete. Like Ludwig Boltzmann faced with the problem of deriving the
laws of continuum thermodynamics from atomistic models,
Fisher, Haldane and Wright developed a statistical theory of
evolution to explain how random mutational events occurring 
in single individuals result in deterministic adaptive changes on the
level of populations. Not surprisingly, then, 
statistical physicists always have been, and are now increasingly attracted to
the study of evolutionary phenomena in biology (see e.g. 
\cite{Drossel2001,Laessig2002,Fisher2007,Blythe2007}). 

In this article we focus on a specific, rather elementary question in 
the mathematical theory of evolution, which was posed in the early
days of the field and remains only partly understood even today:
We ask how rapidly an asexually reproducing, large population 
adapts to a novel environment by generating and incorporating
beneficial mutations. The question originates in the context of the 
Fisher-Muller hypothesis for the evolutionary advantage of sexual vs.
asexual reproduction. Fisher \cite{F1930} and H.J. Muller
\cite{M1932} pointed out that a disadvantage for asexual reproduction 
would arise in populations that are sufficiently large to 
simultaneously accommodate several clones of beneficial mutants.
In the absence of sexual recombination, two beneficial mutations  
that have appeared in different individuals can be combined into 
a single genome only if the second mutation occurs in the offspring 
of the first mutant. This places a limit on the speed with which
the population fitness increases in the asexual population.

The first quantitative treatment of the Fisher-Muller effect was presented 
by Crow and Kimura \cite{CK1965} for a model in which all beneficial
mutations are assumed to have the same effect on the fitness of the
individuals. They arrived at an expression for the speed of evolution
in asexuals which saturates to a finite value in the limit of large 
population size $N \to \infty$, whereas for sexual populations 
the speed increases proportional to $N$. This conclusion was
challenged by Maynard Smith \cite{MS1968}, who showed (for a model
with only two possible mutations) that recombination has no effect on the
speed of adaptation in an infinite population. The resolution of the 
controversy \cite{CK1969,MS1971,Felsenstein1974} made it clear that
the Fisher-Muller effect operates in large, but not in infinite
populations; a first indication of the rather subtle role of
population size, which will be a recurrent theme throughout this article.

Prompted by progress in experimental evolution studies with microbial
populations \cite{dVZGBL1999,RVG2002,EL2003,dVR2006,Hegreness2006,Perfeito2007}, the question of the speed of 
evolution in the setting of Crow and Kimura has been 
reconsidered by several authors in
recent years
\cite{Rouzine2003,Desai2007,DF2007,Beeren2007,Brunet2008,Rouzine2008,Yu2007,Yu2008}. 
Using a variety of approaches, they show that,
rather than approaching a limit for large $N$, the speed grows as 
$\ln N$ in the regime of practical interest, reflecting the increasing
spread of the population distribution along the fitness axis. 
Considerable efforts have been devoted to deriving accurate expressions for prefactors
and sub-asymptotic corrections. 
At the same time more complex models that allow for a distribution of
mutational effects have been introduced and analyzed \cite{GL1998,W2004,PK2007,Fogle2008}.    

The purpose of this article is to review these developments on a level
that is accessible to statistical physicists with no prior knowledge of
population genetics. In the next section we therefore begin by introducing the
basic concepts and models, primarily the discrete time Wright-Fisher model with 
mutations and selection. Section \ref{Sec:Inf_Cal} is devoted to the dynamics of 
an infinitely large population. In this limit the dynamics becomes deterministic
and can be solved exactly using generating function techniques. Although (as 
we will show) real populations operate very far from this limit, the infinite
population behavior serves as a benchmark for the comparison with approximate results
for finite populations, and it yields the important insight that a large population
can be described as a traveling wave in fitness space \cite{Tsimring1996,Rouzine2003}. 
In Section \ref{Sec:Finite} we review the main approaches to the finite population problem.
We provide simple derivations of reasonably accurate expressions for the 
speed of evolution, both for the case of a single type of beneficial
mutations and for models with a distribution of mutational effects,
which are compared to stochastic simulations over a wide range of population sizes.
A preliminary view of the relationship between the two types of models is presented.   
Finally, in Section \ref{Sec:Out} we summarize the article and
discuss some related topics which
point to possible directions for future research.  

\section{\label{Sec:model}Models}

This section introduces the basic concepts and models studied in this paper. 
Models of evolving populations are based on three main features: 
reproduction with inheritance, natural selection, 
and mutation\footnote{Other important features this paper does not 
consider are migration and genetic recombination.}. 
We describe each of these features from the point of 
view of stochastic processes in discrete time. 
For ease of explanation, our description begins with the branching process
well-known in the statistical physics community.

\subsection{Wright-Fisher Model}
We consider here only asexual reproduction that is described by the
number of offspring that each individual produces. This number is
different from one individual to another, depends on many external
events, and is thus described by a random variable. 
In the discrete time branching process without selection,
an individual at time (or generation) $t$ is replaced by
$n_{t+1}$ individuals at time $t+1$ where $n_{t+1}$ is distributed
according to a law $p(n)$ that is the same for all 
individuals\footnote{The fact that all individuals have the same
distribution law implies the absence of natural selection.} and is
constant in time.
The probability $p(0)$ can be seen as the death
probability since the lineage of the individual disappears. The
population is then completely described by its total size $N_t$. This
stochastic process is known as the Bienaym\'e-Galton-Watson process and
describes the growth and the death of a population without restriction
on the size. Simple computations
shows that the average size grows as $\mathbb{E}(N_t) \propto
{\bar n}^t$ where ${\bar n}=\sum_n n p(n)$ is the average number of
children of one individual~\cite{FellerI}. 
This simple system exhibits
a transition to an absorbing (extinct) state as
$\bar n$ varies. When $\bar n \le 1$, the extinction will
occur with probability 1. 
On the other hand, if $\bar{n}>1$, the
population grows exponentially with a finite probability. 
However, such a growth is not realistic because of limitations of the amount of food or resources in the environment. 

In order to take this saturation effect (or environmental capacity) 
into account, we demand 
that the size of population remains constant, with a given value $N$. 
For generality, we now assume the mean number of offspring for each individual
$i$ ($1\le i\le N$) to be $w_i$ in the unrestricted growth case 
described above, and we allow the $w_i$ to be different from each other. 
To make the discussion concrete, we choose a Poisson distribution for the number of
offspring of individual $i$, $p_i(n_i)=w_i^{n_i} \e^{-w_i}/{n_i}!$.
The reproduction mechanism at constant population size can then be modeled 
by conditioning the total number
of offspring $M \equiv \sum_{i} n_i$ to be equal to $N$. 
The joint probability of the $n_i$ (without restriction) is given by
\begin{equation}
\prod_{i=1}^N
p_i(n_i)= \e^{-N\bar w}\prod_{i=1}^N \frac{w_i^{n_i} }{n_i!},
\end{equation}
where $\bar w \equiv \sum_i w_i / N$ is the mean number of offspring per
individual. The probability of observing $M=N$ is 
\begin{equation}
\proba{}{N} = \frac{(N \bar w)^N}{N!} \e^{-N \bar w},
\end{equation}
and, accordingly, the conditioned probability is given by ($\delta$ is the
Kronecker delta symbol)
\begin{equation}
\label{eq:wright-fisher-process}
p(n_1,\ldots,n_N |N) = 
\frac{\delta_{MN}p(n_1)\ldots p(n_N)}{\proba{}{N}} = \delta_{MN}\frac{N!}{n_1!\ldots n_N!} 
\prod_{i}\left (\frac{w_i}{N \bar w} \right )^{n_i},
\end{equation}
\begin{figure}
\centering
\includegraphics[width=0.7\textwidth]{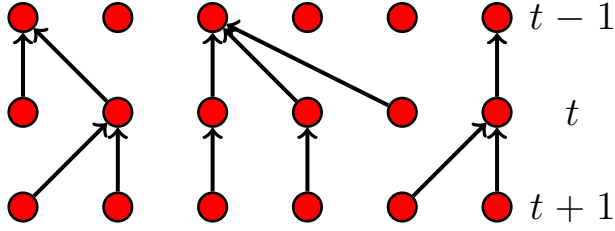}
\caption{\label{Fig:WF}A cartoon illustrating the Wright-Fisher model for a 
population of size $N=6$ over three generations. The arrows indicate how
an individual `chooses' its parent.}
\end{figure}
which is the widely-used Wright-Fisher (WF) model~\cite{W1931,F1930}. 
It becomes then equivalent to the following process: at time $t+1$, each 
individual `chooses'\footnote{In reality, of course, a child cannot choose its 
parent, but this usage of the terminology has no mathematical ambiguity
and is widely used in the literature.} its parent $i$ at time $t$
with probability $w_i/(N\bar w)$; see Fig.~\ref{Fig:WF} as an illustration.
It is obvious that this scheme is not affected by multiplying all $w_i$
by a common factor.
Inheritance is modeled by conferring 
to an offspring the same value of $w_i$ as its parent.

However, the inherited genetic material may go through 
copying errors (or mutations), 
which can result in a child's having different characters 
from its parent. In order to take into account the effects of mutations, 
it is necessary to describe the characteristics of each individual. 
Individuals are usually characterized by a set of parameters, 
the \textit{type} (either the \textit{phenotype}, that describes 
their biological functions and their interactions with their environment, or 
the \textit{genotype}, that specifies their heritable genetic
material). A type is transmitted from the parent to the children up to 
some changes due to genetic mutations. 
For our purposes, the most important characteristic of an individual is 
its \textit{fitness} defined as the average expected number of offspring of 
this individual (even if, for a given realization of the process, the 
effective number of offspring can be different because of
the environmental capacity) in the whole population. In 
the reproduction scheme described above the absolute fitness of 
individual $i$ is thus given by $w_i$, the relative fitness by 
$\chi_i = w_i/\bar w$, and the probability that individual $i$ is chosen 
as a parent in the WF-model is $\chi_i/N$. Fitness differences in the 
population imply selection: individuals with large fitnesses tend to 
generate larger fractions of the populations whereas lineages with small 
fitnesses tend to disappear quickly. 
We return to the question of how fitness is assigned to individuals
below in Section~\ref{Sec:landscape}.

In the language of statistical physics, the WF model as 
defined above may be seen as a mean-field model, 
because it does not take into account any spatial structure of the population: 
any individual can be the parent of any other, without any consideration of 
distance.  This assumption is however realistic if one considers the mixing of 
real populations in a not-so-large environment. The role of spatial structures
in evolution has also been studied for simple models, such 
as the island model~\cite{W1943} and the stepping stone model~\cite{K1953} 
which incorporate migration. 
The present paper focuses on mean-field reproduction models.

The WF model assumes a complete replacement of the population by children
in one generation, i.e. generations do not overlap. 
A model with overlapping generations may be defined by splitting 
the replacement of the population over a longer time. 
A frequently used model that includes overlapping generations as well
as a limited environmental capacity was introduced by 
Moran~\cite{M1958}: at each time step, one individual chosen at random is 
killed and replaced by the child of another individual chosen with probability 
$\chi_i/N$. The time in the Moran model is still discrete, although
the dynamics is evidently close to a scheme where single
individuals are replaced in \textit{continuous} time with rates proportional 
to the $\chi_i$.

Both WF and Moran models have advantages and disadvantages.
Unlike the WF model, the Moran model is amenable to some exact analysis, see
Section~\ref{Sec:fix} for an example.
However, with regard to computational efficiency, the WF model is 
superior to the Moran model when simulating large populations. 
Since the conclusions relevant to biology 
are mostly insensitive to model details, we will base 
our discussion on the discrete time WF model, and comment on the 
corresponding continuous time or Moran model where appropriate.

\subsection{\label{Sec:landscape}Fitness landscapes and selection coefficients}

The main difficulty in modelling biological evolution within the
framework described so far is the choice of the functional relationship $w(\mathcal{C})$
between the type $\mathcal{C}$ of an individual and its fitness, referred to as the 
\textit{fitness landscape}, which encodes in a single
parameter the complex interactions of a type with its environment~\cite{JK2006}. At least two distinct approaches circumvent this difficulty: one can either 
try to measure the function $w(\mathcal{C})$ from experimental data if the 
set of types is reduced~\cite{deVisser2009}, or choose the fitness landscape 
at random from some suitable ensemble. In the last case, a widely-used further 
simplification consists in describing the individuals only by their fitnesses 
and ignoring the underlying structure of the types $\mathcal{C}$; mutations 
are then described only by changing the fitness of an individual by a random 
amount. This can be justified if the number of types is very large,
so that every mutation effectively generates a new type that has never appeared before in the population. In population 
genetics this is known as the infinite number of sites 
approach~\cite{K1969,Park2008},
and it will be used throughout this paper. 
 
Each offspring has a probability $U$ per generation of acquiring a mutation 
and this mutation changes the parental fitness $w_i$ to the fitness
$w_i'$ of the offspring.
In this paper, mutations are assumed to act multiplicatively on the fitness 
\(w_i\) and so the fitness $w'_i$ after mutation is given by
\begin{equation}
\label{Eq:multiplicative} 
w'_i=w_i (1+s) 
\end{equation}
where the \textit{selection coefficient} $s$ is a random 
variable with a distribution $g(s)$. Mutations with $s > 0$ are 
\textit{beneficial} and those with $s < 0$ \textit{deleterious}. 
Recall that if all the $w_i$ are multiplied by the same quantity, then 
the relative fitnesses $\chi_i$ do not change, 
which justifies the multiplicative action of the mutations\footnote{An alternative
scheme where the mutant fitness $w_i'$ itself is chosen at random
was investigated in \cite{Park2008}, see Section~\ref{Sec:Out} for further
discussion.}~\cite{JK2006}. 
One expects the relative fitnesses to reach a stationary distribution at 
long times such that the average fitness $\bar{w}(t)$ will increase 
(or decrease) exponentially with a rate referred to as the \textit{speed of evolution}
\begin{equation}
\label{Eq:speed_def}
 v_N= \lim_{t\to\infty} \frac{\langle \ln \bar{w}(t)\rangle}{t},
\end{equation}
where the angular brackets denote an average over all realizations.
This speed depends on the population size $N$ as well as on the mutation rate 
$U$ and on the distribution $g(s)$ of the mutation. Two main contributions sum 
up to give the speed $v_N$: the change of mean fitness due to mutations and 
the selection pressure that selects individuals with larger $w_i$. For the 
WF model, these contributions are made explicit through a result 
obtained by Guess~\cite{G1974A,G1974T}:
\begin{equation}
 \label{eq:guessrelation}
v_N = U \int \; \ln(1+s) g(s) ds + \left\langle \frac{1}{N} \sum_{i=1}^N (\chi_i-1)\ln \chi_i \right\rangle_\text{stat},
\end{equation}
where $\langle \cdot \rangle_\text{stat}$ indicates an average over the stationary 
measure of the $\chi_i$. Only the second term of this expression 
(which is always nonnegative) 
is related to selection. However it 
is also the difficult part to study since the stationary distribution of 
the $\chi_i$ is generally unknown and hard to compute. 
If the distribution of relative fitness $\chi_i$ is concentrated around $1$,
the second term can be approximated by the variance of the distribution of
relative fitness.
This result is reminiscent of Fisher's fundamental theorem~\cite{F1930} 
which states that the speed of evolution is proportional to the variance 
of the fitness distribution.

In the present paper the dependence of the speed of evolution on the distribution $g(s)$ of selection 
coefficients is a central theme. As it turns out that deleterious mutations
do not affect the adaptation of large populations when at least 
some fraction of mutations is beneficial\footnote{When all mutations are deleterious, the fitness decreases at constant speed and
the problem is known as Muller's ratchet, see \cite{Rouzine2003,Rouzine2008,Yu2008,Jain2008} and references therein.}, 
only beneficial mutations ($s > 0$) will be considered
in the following. The mutation rate (per generation) $U$ then refers to the rate of \textit{beneficial mutations}, which 
is exceedingly small in natural populations: 
experimental estimates for bacteria range from $10^{-7}$ to $10^{-4}$~\cite{Hegreness2006,Perfeito2007}. 
The distribution of selection coefficients of beneficial mutations is very    
difficult to determine experimentally, and the choice of a realistic form remains an open
question~\cite{Eyre2007}. Moreover, the experimental determination of evolutionary parameters
such as the mutation rate $U$ and the typical size of selection coefficients depends strongly
on the assumptions made about the shape of $g(s)$~\cite{Hegreness2006}.

It has been argued that, because viable populations are already
well adapted to their environment, fitness coefficients associated with beneficial mutations
occur in the extreme high fitness tail of the underlying `bare' fitness distribution,
and therefore the shape of $g(s)$ should be given by one of the
invariant distributions of extreme value statistics \cite{O2003,Joyce2008}. 
Here we will consider two choices for this distribution. 
The first one (\textit{model I}) describes the situation where all mutations have the same selective strength $s_b$,
\begin{equation}
\label{gdelta}
g^{(\infty)}(s) = \delta(s - s_b).
\end{equation}
The second class of distributions (\textit{model II}) is supported on the whole non-negative real axis and 
decays as a stretched exponential \cite{DF2007,Fogle2008},
\begin{equation}
\label{eq:def:gbeta}
\gbeta(s) = (\beta/s_b) (s/s_b)^{\beta-1} \exp( -(s/s_b)^\beta),
\end{equation}
where the factor $(s/s_b)^{\beta-1}$ has been introduced for computational convenience. 
For $\beta=1$, one recovers the widely-used exponential distribution \cite{O2003,GL1998,W2004,PK2007}, 
whereas for $\beta \to \infty$ (\ref{eq:def:gbeta}) reduces to (\ref{gdelta}). We note for later
reference that the mean of $\gbeta$ is $\Gamma(1+1/\beta) s_b$. Typical values of selection coefficients
obtained from evolution experiments with bacteria lie in the range $s_b \approx 0.01 - 0.05$ \cite{Hegreness2006,Perfeito2007}.
Thus both $U$ and $s_b$ can be treated as small parameters, with $U \ll s_b$, 
in most of what follows.

\section{\label{Sec:Inf_Cal}Infinite population dynamics}

This section studies the WF model in the infinite population limit
which is described by a deterministic evolution equation.
Some of the material of this section is also found in the online supporting
information of \cite{PK2007}. Since it was shown in \cite{PK2007} that 
deleterious mutations do not contribute to the speed in the infinite 
population limit, all mutations are assumed to be beneficial in the following.
The model will first be solved using a discrete set of fitness values, and the 
transition to a continuous fitness space will be performed in 
Section~\ref{Continuum}. 
 
\subsection{\label{Sec:Inf_formal}The evolution equation and its formal solution}
Let $f_{t}(n,k)$ denote the frequency of 
individuals with $n$ (beneficial) mutations and
with fitness $\e^{k s_0}$ at generation $t$; here $s_0>0$
and $k$ is a non-negative integer. Note that 
$f_t(n,k)$ does not discern different types
which have the same number of mutations and the same fitness.
The restriction to fitnesses $\geq 1$
is irrelevant due to the invariance of the dynamics under multiplication 
of absolute fitnesses by a common factor. 
The mean fitness of the population at generation $t$ is 
\begin{equation}
\bar w(t) = \sum_{n,k} \e^{k s_0} f_{t}(n,k).
\end{equation}
If there are no mutations, the frequency at the next generation is given by
\begin{equation}
\tilde f_{t+1}(n,k) = \frac{1}{\bar w(t)} \e^{k s_0}
 f_t(n,k),
\label{Eq:WF_sel}
\end{equation}
which is equal to the expected frequency at generation $t+1$ for a finite
population.

After reproduction, mutations can change the type of the offspring.
With probability $U$, mutations hit
an individual and with probability $1 - U$ 
the offspring keeps the type inherited from its parent.
For simplicity, we assume that a single mutation occurs in 
a single mutation event (see \cite{PK2007} for more general cases).
For each mutation a positive integer from a distribution 
$g_0(l)$ with strictly positive $l$ 
is drawn and then the fitness of the offspring is that of 
its parent multiplied by $\e^{l s_0}$.
It is convenient to introduce the generating function of $g_0(l)$,
\begin{equation}
G(z) = \sum_{l=1}^\infty z^l g_0(l),
\end{equation}
with the normalization $G(1) = 1$.
Including the effect of mutations along with the selection
step in Eq.~\eqref{Eq:WF_sel}, the frequency change becomes
\begin{equation}
f_{t+1}(n,k) = (1 - U) \tilde f_{t+1}(n,k)
+ U\sum_{l=1}^{k} \tilde f_{t+1} (n-1,k-l) 
g_0(l),
\label{Eq:WF}
\end{equation}
which is the main equation to be analyzed in this section.

The generating function of the frequency
\begin{equation}
F_t(\xi,z) = \sum_{n,k} \xi^n z^k f_t(n,k)
\end{equation}
satisfies
\begin{equation}
 F_{t+1}\left (\xi,z\right )
= \frac{F_t\left (\xi,z \e^{s_0}\right )}{F_t\left (1,\e^{s_0}\right )}
\left [ 1 - U + U \xi G\left (z\right ) \right ],
\label{Eq:main_eq}
\end{equation}
where we have used the relations 
\begin{eqnarray}
&&\sum_{n,k} \xi^n z^k \tilde f_t(n,k)=
\frac{F_t(\xi,z \e^{s_0})}{F_t(1,\e^{s_0})},\\
&&\bar w(t) = F_t(1,\e^{s_0}),
\label{Eq:MW}
\end{eqnarray}
and the property of the convolution. 
Iterating Eq.~\eqref{Eq:main_eq} backwards until the initial time gives
\begin{equation}
\begin{aligned}
F_t\left (\xi,z\right ) 
=\frac{F_0\left (\xi,z \e^{s_0t} \right )}{F_0\left (1,\e^{s_0t} \right )}
\prod_{\tau=0}^{t-1} \frac{1+ 
u \xi G\left (\e^{s_0\tau} z\right ) }{1+ u 
G\left (\e^{s_0\tau}\right ) },
\end{aligned}
\label{Eq:gen_sol}
\end{equation}
where $u = U/(1 - U)$.
One can check that Eq.~\eqref{Eq:gen_sol} solves Eq.~\eqref{Eq:main_eq} by 
substitution. 

\subsection{General asymptotic behavior}
Using Eqs.~\eqref{Eq:MW} and \eqref{Eq:gen_sol}, 
the mean fitness at generation $t$ becomes
\begin{equation}
\ln \bar w(t) = \ln \frac{F_0\left (1,\e^{s_0 (t+1)}\right )}{
F_0\left (1,\e^{s_0 t} \right )} +
\ln \left (1- U + U  G \left (\e^{s_0 t} \right) \right ).
\label{Eq:meanfitness}
\end{equation}
If initially there is a finite $K_0$ such that 
$f_{0}(n,k) = 0$ for $k>K_0$, the first term arising from the initial condition
saturates and does not contribute to the speed in the long time limit.
On the other hand, if such a $K_0$ does not exit, 
the initial condition can affect the fitness increase indefinitely.
For example, let $f_{0}(n,k) = \delta_{n,0} \e^{-\eta} \eta^k
/k!$ with the generating function $F_0(\xi ,z) = \e^{\eta(
z - 1)}$. The first term on the right hand side of 
Eq.~\eqref{Eq:meanfitness} 
then becomes $\eta (\e^{s_0}-1) \e^{s_0t}$
which does not allow a finite
increase rate even in the absence of mutations. This is a peculiarity of the
selection dynamics in the infinite population limit and it is not difficult
to understand why this happens. 
Since the selection confers exponential growth to all types with 
fitness larger than the average $\bar w(t)$ and there are always individuals of such
types at any generation $t$ due to the unbounded initial condition, 
the mean fitness can grow indefinitely without
recourse to beneficial mutations. Because this is a rather artificial situation
which has no biological relevance, we assume the existence of $K_0$ 
in what follows. Actually, for simplicity 
the initial condition 
\begin{equation}
\label{initial}
f_0(n,k) = \delta_{n0}\delta_{k0}, \;\;\;F_0(z,\xi) = 1 
\end{equation}
will be used throughout this paper.

As $t \rightarrow \infty$, the speed is determined solely by the generating
function of beneficial mutations. 
Let $K_{\max} = \max_k \{k: g_0(k)\neq 0\}$, then due to the exponential growth
of the argument of $G(\e^{s_0 t})$ the speed for the infinite size population
becomes
\begin{equation} 
v_\infty
= K_{\max}  s_0.
\label{Eq:Det_speed}
\end{equation}
This shows that the mutation of largest effect governs the speed, which is
not surprising because genetic drift (a term referring to the stochastic 
loss of a beneficial mutation in a finite population, see Section \ref{Sec:fix}) 
is not operative.
If we take $K_{max} \rightarrow \infty$ with $s_0$ fixed, the speed
diverges.
Note that the speed in the infinite population limit does not depend on the
mutation rate. It is only determined by the maximum value 
of the fitness increase by a single beneficial mutation event.

Another peculiarity of the infinite population limit is the possibility that
the fitness becomes infinite at finite time.
If $G(z)$ is not an entire function, 
the series defining the generating function has a finite radius of convergence, say ${\cal R}$,
beyond which the series diverges. Hence when $\e^{s_0 t} > {\cal R}$ or
$t > \ln {\cal R} / s_0$, the mean fitness becomes infinite. For example, let $g(l) = (1-p) p^{l-1}$ which yield
$G(z) = (1-p) z /( 1 - p z)$ for $p z < 1$ and infinite otherwise. 
Hence for $ t > - \ln p /s_0$, the fitness becomes infinite.
The radius of convergence for this example is ${\cal R} = 1/p$.
Also note that the radius of convergence cannot be smaller than 1 because
the generating function of probability is absolutely convergent 
for $|z|\le 1$ by definition.
In the following, $G(z)$ is assumed to be an entire function, that is,
$g_0(l)$ is assumed to decay faster than exponential in the asymptotic
regime.

We now proceed to calculate the mean and 
variance of the number of accumulated mutations in the infinite 
population limit. 
First, the mean number of mutations is calculated as
\begin{equation}
 \bar n(t) = \left .  \frac{\partial}{\partial \xi} \ln F_t(\xi,1) \right |_{\xi=1}
= t - \sum_{\tau = 0}^{t-1}  \frac{1}{
1 + u G(\e^{s_0\tau}) }.
\label{Eq:mean_bensol}
\end{equation}
Since $G(e^{s_0 \tau})$ grows at least exponentially
with $\tau$, the second term approaches a finite value.
Clearly Eq.~\eqref{Eq:mean_bensol} gives the large population limit of the 
\textit{substitution rate} $k$, defined here as the infinite time limit of $\bar n(t) /t$:
\begin{equation}
k = \lim_{t \rightarrow \infty} \frac{\bar n(t)}{t} = 
1.
\label{Eq:rate_sub}
\end{equation}
The variance of the number of mutations reads
\begin{eqnarray}
\delta n(t)^2 
= \left . \left ( \xi \frac{\partial}{\partial \xi} \right )^2
\ln F_t(\xi,1) \right |_{\xi=1}=
 \sum_{\tau = 0}^{t-1} \frac{u  
  G(\e^{s_0 \tau})  }{
[1 + u G(\e^{s_0 \tau})]^2},
\label{Eq:var_ben}
\end{eqnarray}
which has finite limit as $t\rightarrow \infty$.

\subsection{\label{Sec:Inf_sol} Case studies}
Using the results presented above, we study the detailed evolution
for two specific examples. To begin with, Sec.~\ref{Sec:GLinf} 
studies the simple case that $g_0(l) = \delta_{l1}$ which corresponds to 
(\ref{gdelta}) with $s_b = s_0$. Then in Sec.~\ref{Continuum} we generalize
our solution to a continuous fitness distribution such as (\ref{eq:def:gbeta}).
 
\subsubsection{\label{Sec:GLinf}The case of a single selection coefficient}
When $g_0(l) = \delta_{l1}$, the calculation is rather straightforward.
Because the number of mutations fully specifies the fitness of a 
type, we replace $f_t(n,n)$ by $f_t(n)$ throughout this subsection.
From Eq.~\eqref{Eq:meanfitness}, the mean fitness becomes
\begin{equation}
\ln \bar w(t) =\ln \left (  1 - U + U \e^{s_0 t}\right ) \approx s_0 
\left (t - \frac{1}{s_0} \ln \frac{1}{U} \right )
= s_0 (t-t_0)
\label{Eq:Infmean}
\end{equation}
with
\begin{equation}
\label{t0}
t_0 = \frac{1}{s_0} \ln \frac{1}{U},
\end{equation}
which gives $v_\infty = s_0$. 
The mean number of mutations in the long time limit can be calculated
from Eq.~\eqref{Eq:mean_bensol} as
\begin{equation}
\bar n(t) \rightarrow t - \sum_{\tau=0}^\infty
\frac{1}{1+u \e^{s_0 \tau}}
\approx t - \int_0^\infty d\tau \frac{1}{1+u \e^{s_0\tau}}
= t - \frac{1}{s_0} \ln \frac{1+U}{U} \approx t - t_0,
\label{Eq:Infnum}
\end{equation}
where we approximate the summation by an integral assuming $s_0 \ll 1$,
and $U \ll 1$.
Not surprisingly, $s_0 \bar n(t) \approx \ln \bar w(t)$ in the
long time limit.
Likewise, the variance of the number of mutations is calculated as
\begin{eqnarray}
\delta n(t)^2 \rightarrow \sum_{\tau=0}^\infty
\frac{u \e^{s_0 \tau}}{\left ( 1 + u \e^{s_0 \tau} \right)^2}
\approx \int_0^\infty d \tau \frac{u \e^{s_0 \tau}}{\left ( 1 + u \e^{s_0 \tau} \right)^2} = \frac{1-U}{s_0}.
\label{Eq:Infdev}
\end{eqnarray}

Now we will show that the frequency distribution in the asymptotic
limit can be approximated by a Gaussian.
From Eq.~\eqref{Eq:WF} with $\bar w \approx \e^{s_0(t-t_0)}$,
the frequency at generation $t$ can be approximated as
\begin{eqnarray}
f_{t}(n) 
&\approx& f_{t-1}(n) \e^{s_0(n-(t-1)+t_0)} \nonumber\\
&\approx& f_{n}(n) \exp\left (s_0\sum_{\tau=1}^{t-n} 
\left ( n + t_0 - t + \tau \right )\right ) 
\approx
f_{n}(n)\e^{s_0 t_0^2/2} \e^{-s_0(n -t + t_0)^2/2},
\label{Eq:gauss_travel}
\end{eqnarray}
where $n$ and $t$ are assumed sufficiently large and we neglect the
effect of mutations. 
Next we show that $f_n(n) \approx \e^{-s_0 t_0^2/2}$ at long times,
which concludes the demonstration that $f_t(n)$ becomes Gaussian.
Under the assumptions of our model the largest number of mutations accumulated
by an individual up to $t$ is $t$, and 
from Eq.~\eqref{Eq:gen_sol} the 
frequency of such individuals is
\begin{equation}
f_t(t) = \prod_{\tau=0}^{t-1} \frac{u \e^{s_0 \tau}}{1 + u \e^{s_0 \tau}}.
\label{Eq:largest_mut}
\end{equation}
Since $u \e^{s_0 t}$ becomes larger than 1 at 
$t \approx t_0$, the term
$u \e^{s_0 \tau}$ in the denominator of Eq.~\eqref{Eq:largest_mut} 
makes a dominant (negligible) contribution for $t> t_0$ ($t<t_0$).
Thus, we may approximate Eq.~\eqref{Eq:largest_mut} in the long time limit as
\begin{equation}
\lim_{t\rightarrow \infty} f_t(t) \approx U^{t_0} \e^{s_0 t_0 (t_0 -1)/2}
 \approx \e^{-s_0 t_0^2/2},
\label{Eq:edge_frequency}
\end{equation}
which shows that $f_t(n)$ is well described by
a travelling wave in the form of Gaussian.

The above consideration gives an interesting criterion for
the population size beyond which the infinite population 
dynamics becomes valid. If the population size is larger than
\begin{equation}
\label{Nc}
N_c \equiv \exp(s_0 t_0^2/2) = \exp[\ln^2 U/(2s_0)], 
\end{equation}
the number of fittest individuals 
at a given generation is not smaller than 1 for all times $t$
(note that $f_t(t)$ is a decreasing function of $t$). Since
the selection coefficient of the types with $t$ mutations compared to the
mean fitness is approximately $\e^{s_0 t} / \bar w(t) -1 \approx 1/U \gg 1$,
we can neglect the possible loss of such a type 
by genetic drift even if it is rare, which means that the infinite population
dynamics describes a finite population with $N\ge N_c$.
To provide an impression of how large $N_c$ is, we  choose typical values
$s_0 = 0.02$ and $U = 10^{-5}$, which gives\footnote{The more accurate value obtained by
exact numerical calculation is $N_c \approx 2\times 10^{1477}$.} 
$N_c \approx 10^{1439}$. 

\begin{figure}
\centerline{\includegraphics[width=1.0\textwidth]{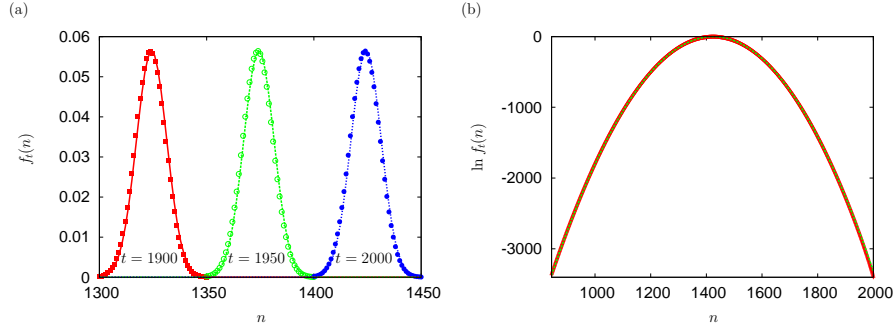}}
\caption{\label{Fig:fre_inf} (a) Frequency distribution of the infinite
population dynamics for the case of $g_0(l) = \delta_{l1}$ with
$s_0 = 0.02$ and $U=10^{-5}$. The distributions
are shown at $t=1900$ (left), $t=1950$ (middle), and $t=2000$ (right).
The peak is located at $t + (\ln U)/s_0 \approx t - 575.65$.
(b) Plot of $\ln f_t(n)$ at $t=2000$ as a function of $n$ in
comparison to Eq.~\eqref{Eq:inf_gauss}. Only a tiny deviation around
$n=2000$ is visible.
}
\end{figure}

To include the effect of mutations,
we use Eqs.~\eqref{Eq:Infnum} and \eqref{Eq:Infdev} to write
the frequency distribution in the form
\begin{equation}
f_t(n) \approx \frac{1}{\sqrt{2 \pi (1-U)/s_0}} 
\exp\left (-\frac{(n-\bar n(t))^2}{2 (1-U)/s_0} \right ),
\label{Eq:inf_gauss}
\end{equation}
where the prefactor is fixed by normalization.
Note that for sufficiently small $U$, Eq.~\eqref{Eq:gauss_travel} 
is consistent with Eq.~\eqref{Eq:inf_gauss}. 
Figure~\ref{Fig:fre_inf} compares the numerically obtained frequency
distribution with Eq.~\eqref{Eq:inf_gauss} for $U=10^{-5}$ and
$s_0 = 0.02$.

The idea that evolution can be described as a traveling wave moving at constant
speed along the fitness axis was first presented by 
Tsimring et al.\cite{Tsimring1996,Kessler1997},
who considered the continuous time version of the model with multiplicative
mutations and a single selection coefficient. In the continuous time case the 
speed of evolution diverges in the infinite population limit, because there
is no bound on the number of mutations that a single individual can accumulate
in a given time. A wave moving at finite speed is obtained only if the finite
size of the population is introduced at least on the level of a lower cutoff on 
the frequency distribution.

\subsubsection{\label{Continuum}Solution for a continuous fitness space}
In this section, we explain how the above calculation can be generalized
to a continuous fitness space.
We will use Eq.~\eqref{eq:def:gbeta} for the distribution of
selection coefficients. To connect to the results in Sect.~\ref{Sec:Inf_formal},
we perform a change of variables such that $\e^x = 1+s$, where $x$ denotes
the continuous version of $k s_0$. When
$s$ is drawn from $\gbeta (s)$, the probability density for $x$ becomes
\begin{equation}
\gbeta_0 (x) =  
\beta \left (
\frac{\e^{x} - 1}{s_b} \right )^{\beta-1} 
\exp\left (-\left (\frac{\e^{x} - 1}{s_b}
\right )^\beta \right ) \e^{ x} \frac{1}{s_b}.
\end{equation}
Setting $x = l s_0$, the corresponding discrete distribution is
\begin{equation}
g_0(l) = {\cal N}(s_0)\beta \left (
\frac{\e^{s_0 l} - 1}{s_b} \right )^{\beta-1} 
\exp\left (-\left (\frac{\e^{l s_0} - 1}{s_b}
\right )^\beta \right ) \e^{ l s_0} \frac{s_0}{s_b},
\end{equation}
where ${\cal N}(s_0)$ is the normalization constant which approaches  
$1$ as $s_0 \rightarrow 0$. We now follow Sect.~\ref{Sec:Inf_formal},
and calculate $G(\e^{s_0 t})$ as
\begin{eqnarray}
&&G(\e^{s_0 t}) = {\cal N}(s_0)
\sum_{k=1}^\infty \beta \left (
\frac{\e^{s_0 k} - 1}{s_b} \right )^{\beta-1}
\exp\left (-\left (\frac{\e^{k s_0} - 1}{s_b}
\right )^\beta \right ) \e^{ k s_0(t+1)} \frac{s_0}{s_b}
\nonumber
\\
&&\longrightarrow  
\int_0^\infty dx \left (
\frac{\e^{x} - 1}{s_b} \right )^{\beta-1}
\exp\left (-\left( \frac{\e^x - 1}{s_b} \right )^\beta \right ) 
\e^{x(t+1)} \frac{\beta}{s_b} dx ,
\end{eqnarray} 
where we take $s_0 \rightarrow 0$ with $s_0 k = x$ finite.
Letting $y^{1/\beta} = (\e^{x} - 1)/s_b$, the above integral, say $W_t$, becomes
\begin{equation}
W_t = \int_0^\infty (1 + s_b y^{1/\beta})^t e^{-y} 
dy 
= \int_0^\infty \exp\left (-y+ t \ln (1+s_b y^{1/\beta})\right ) dy.
\label{Eq:Wt}
\end{equation}
Using the steepest descent method, this can be
approximated as
\begin{equation}
W_t \sim \exp\left (-y_c + t \ln \left (1+  s_b y_c^{1/\beta}\right )\right ),
\end{equation}
where $y_c$ is the solution of the saddle point equation
\begin{equation}
y_c^{1-\frac{1}{\beta}} + s_b y_c = \frac{s_b t}{\beta}
\rightarrow y_c \sim \frac{t}{\beta}
\end{equation}
Thus, the leading behavior of $\ln W_t$ becomes
$t\ln t/\beta$, which along with 
Eq.~\eqref{Eq:meanfitness} yields
\begin{equation}
\begin{aligned}
\ln \bar w(t)\sim \ln W_{t} \sim \frac{t \ln t}{\beta}
\label{Eq:Wtapp}
\end{aligned}
\end{equation}
for any finite $\beta$. On the other hand, for $\beta \to \infty$
(\ref{Eq:Wt}) yields $W_t = (1 + s_b)^t$ and hence $\ln \bar{w}(t) \sim t$.

For $\beta = 1$, a more accurate approximation can be found in 
Ref.~\cite{PK2007}.
Here we observe that the speed $\ln \bar w(t) / t$ increases
logarithmically with time irrespective of the value of $\beta$.
The linear relation~\eqref{Eq:mean_bensol} for the mean number of beneficial
mutations is still valid, because the summation on the right hand
side of Eq.~\eqref{Eq:mean_bensol} approaches to
a  non-negative finite number in the continuum limit $s_0 \to 0$. 

We conclude that the speed of evolution is infinite in the infinite
population limit for distributions
of selection coefficients like (\ref{eq:def:gbeta}), which have
unbounded support. Superficially this is reminiscent of the 
situation in the continuous time model with a single selection
coefficient considered in \cite{Tsimring1996,Kessler1997}, but it is important
to note that the reasons for the divergence of the speed are quite
different in the two cases. In the continuous time setting the speed
diverges because the \textit{number} of mutations accumulated in a given
time is unbounded, whereas in the discrete time model the divergence
reflects that a \textit{single} mutation can have an arbitrarily large effect. 
We will encounter a similar dichotomy in the discussion of the finite population
dynamics in the next section.

\section{\label{Sec:Finite}Finite populations}
\subsection{\label{Sec:fix}Genetic drift, fixation and clonal interference}

Consider a single beneficial mutation with selection coefficient $s >
0$ which is introduced into an initially homogeneous population. 
Following the evolution of the population under WF dynamics without
allowing for further mutations ($U=0$),  
one can distinguish different time scales. The 
survival of the mutation during the first few generations is very fragile, due to the
stochasticity of the reproduction process: the number of individuals
carrying the mutation is small and the variance is of the same order
as the mean. These fluctuations are called \textit{genetic drift}. After
this drift phase, either the mutation goes extinct (with some probability
$1-\pi_N(s)$) or the number of individuals becomes large enough 
(with probability $\pi_N(s)$), so that stochastic fluctuations can
then be neglected and the evolution can be considered 
deterministic. 
A mutation that has reached the latter regime is called
\textit{established}\footnote{Although this terminology is mathematically
ambiguous, it is widely used in the community because, we think, 
it is inspirational.}~\cite{MS1971,Felsenstein1974,Fisher2007},
and it will (in the absence of other mutations) eventually take over the entire
population. This process is called \textit{fixation}, $\pi_N(s)$ is
the fixation probability, and 
the time needed for a mutation that survives to spread all over the
population is the fixation time $t_\text{fix}$.

The fixation probability for the Moran 
model is given by~\cite{Durrett2002} 
\begin{equation}
\label{Moranfix}
\pi_N(s) = \frac{s}{1+s - (1+s)^{-(N-1)}},
\end{equation}
but for the Wright-Fisher model only approximate expressions
are available \cite{Sella2005,Barrett2006}.  
A widely used formula is~\cite{K1962}
\begin{equation}
\label{Kfix}
\pi_N(s) = \frac{1 - e^{-2s}}{1 - e^{-2Ns}}.
\end{equation}
For $s \to 0$ both (\ref{Moranfix}) and (\ref{Kfix}) reduce to $\pi_N = 1/N$,
as is obvious from a symmetry argument: When the fitness of the mutant is equal to that
of the background population, the probability of fixation is the same for all $N$ individuals.
Both expressions show that the fixation of deleterious mutations ($s < 0$) is
exponentially suppressed for large $N$, while the fixation probability
for beneficial mutations becomes independent of $N$, reducing for (\ref{Kfix}) 
to 
\begin{equation}
\label{Largefix}
\pi_\infty(s) \equiv \pi(s) = 1 - e^{-2s}.
\end{equation}
When $s$ is small (as will often be the case) this can be further simplified to
\begin{equation}
\label{Smallfix}
 \pi(s) \simeq 2s, 
\end{equation}
while $\pi(s) \simeq s$ for the Moran model.

In the limit $N \to \infty$ the restriction on the size of the growing mutant clone is irrelevant and
the WF-model reduces to~\footnote{The derivation of the WF model from
the branching process in Sect.~\ref{Sec:model} easily explains this connection.} 
a Bienaym\'e-Galton-Watson branching process with a Poisson offspring
distribution of mean $1 + s$. The fixation probability is then equal to the survival probability
of the branching process, which satisfies the implicit relation \cite{Haldane1927,FellerI,Barrett2006}
\begin{equation}
\label{Haldane}
\pi = 1 - e^{-(1+s)\pi}.
\end{equation}
Expanding Eq.~\eqref{Haldane} to second order we recover Eq.~\eqref{Smallfix} for small $s$, but 
for large $s$ the exact fixation probability approaches unity as $1 - \pi \approx e^{-(1+s)}$, in contrast
to Eq.~\eqref{Largefix}. In the following we nevertheless use (\ref{Largefix}) when values of the 
fixation probability are required for the full range of selection coefficients, and (\ref{Smallfix}) when
$s$ is small. 

The approximation by a branching process is also useful in deriving a
heuristic estimate of the population size required for a mutant clone to
become established \cite{MS1971}. In this approximation the average
population size of the clone grows as $(1+s)^t \approx e^{st}$. However, since this
average includes also instances where the clone goes extinct (with
probability $1 - \pi$), the population size conditioned on survival of
the clone is larger by a factor $1/\pi \approx 1/2s$. Such a clone
thus looks as if it started out containing already $\sim 1/2s$
individuals, which is precisely the threshold size separating stochastic
from deterministic growth (see e.g. \cite{DF2007,Brunet2008}
for a detailed treatment of this point).     

In order to get some intuition about the fixation time $t_\text{fix}$, one can look at the deterministic evolution of a mutant of
type $A$ that appears in a population consisting of the ``wild type'' $B$. 
The fitnesses can be taken as $w_A=(1+s)$ and $w_B=1$. We assume that the 
type $A$ has survived genetic drift and we have a frequency $a_t$ of individuals of type $A$ and $b_t=1-a_t$ of individuals of type $B$. 
The deterministic evolution is thus given by
\begin{equation}
\begin{cases}
   a_{t+1} = \frac{1+s}{\bar{w}_t} a_t ,\\
   b_{t+1} = \frac{1}{\bar{w}_t} b_t ,\\
  \bar{w_t} = (1+s)a_t+b_t
\end{cases}
\end{equation}
and the solution is
\begin{equation}
\label{abclone}
 b_t = \frac{b_0}{a_0(1+s)^t + b_0},\qquad a_t=1-b_t.
\end{equation}
For a finite population of large size $N$, the type $B$ can be
considered as extinct when $b_t=1/N$. 
With the initial condition of a single mutant, $a_0 = 1/N$, this expression gives the fixation time for large $N$ as
\begin{equation}
 t_\text{fix} \simeq \frac{2\ln( N-1)}{\ln(1+s)} \simeq \frac{2 \ln N}{s}
\label{Eq:Fix}
\end{equation}
when $s$ is small.

For later purposes, we also need the total number of individuals of
type $B$ that have existed during the fixation time of $A$. We note that
\begin{equation}
a_{t_\text{fix}-t} = \frac{ (1+s)^{t_\text{fix}-t}}{ (1+s)^{t_\text{fix}-t} +N-1} = \frac{N-1}{(1+s)^t + N-1} = b_t
\label{Eq:AB}
\end{equation}
where we have used $(1+s)^{t_\text{fix}}=(N-1)^2$. We can thus
conclude that, 
during the fixation of type $A$, 
one has $\int a_t dt \simeq \int b_t dt$, so that the total number of
individuals of type 
$B$ is $\simeq N t_\text{fix}/2 \simeq N\ln N/\ln(1+s)$.

\begin{figure}
\centerline{\includegraphics[width=0.7\textwidth]{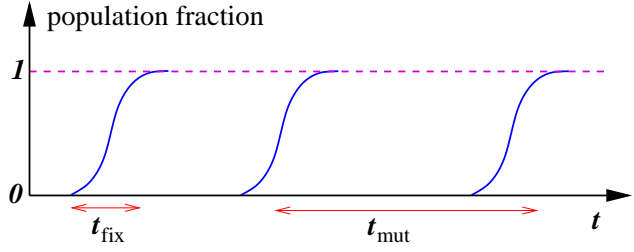}}
\caption{\label{Fig:periodic} In the periodic selection regime 
$t_\text{mut} \gg t_\text{fix}$ and beneficial mutations fix
independently of each other. Each blue line represents a
selective sweep.}
\end{figure}

This simple example shows the dependence of $t_\text{fix}$ on $N$ when
the mutation rate $U$ is set to $0$ after the emergence of the mutant type $A$.
If $U$ is non-zero, the expression for $t_\text{fix}$ is
valid only as long as $U$ is small enough, so that no new mutation
emerges before the fixation of the previous one. The average time
between two mutations that survive genetic drift is
\begin{equation}
\label{tmut}
t_\text{mut} = \frac{1}{N U \pi(s)} \simeq \frac{1}{2 NU s}. 
\end{equation}
If $t_\text{mut} \gg t_\text{fix}$, i.e. if 
\begin{equation}
\label{onset}
N\ln N U \ll 1  
\end{equation}
for $s$ small, then no
mutation interferes and we are in the \textit{periodic selection regime} for which 
\begin{equation}
\label{periodic}
v_N = \frac{s}{t_\text{mut}}\propto s^2 N U.
\end{equation}
This situation is sketched in Fig.~\ref{Fig:periodic}. The main feature of this regime
is that the \textit{selective sweeps} associated with different
beneficial mutations are independent and well separated in time, and therefore the speed
of evolution is directly proportional to the supply of beneficial mutations $N U$.

\begin{figure}
\centerline{\includegraphics[width=0.8\textwidth]{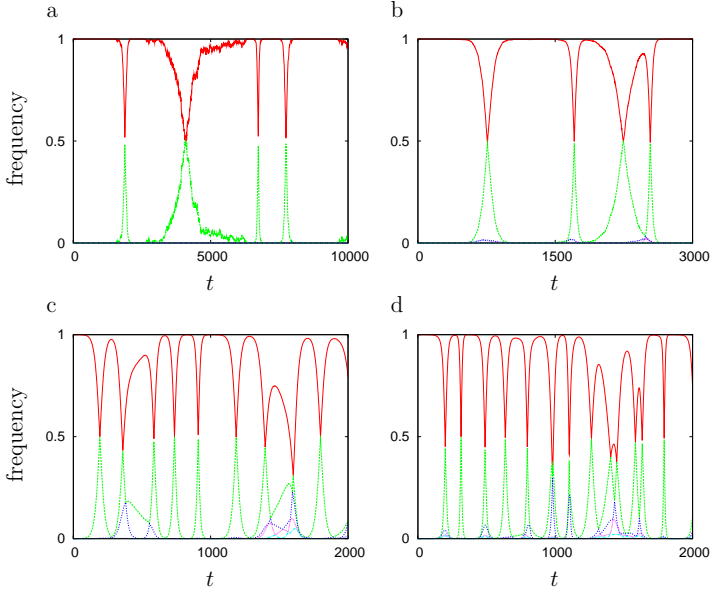}}
\caption{\label{Fig:CI}
The frequencies of the five most populated genotypes are shown in different
colors for the WF model using the distribution \eqref{eq:def:gbeta} with
$\beta=1$, $U=10^{-6}$, $s_b=0.02$. The  population sizes are
(a) $N=10^4$, (b) $N=10^5$, (c) $N=10^6$, and (d) $N=10^7$, respectively.
From $N=10^5$ onward, where $N U \ln N \approx 1.15$, the third most populated
genotype becomes visible and the distinction between $t_\text{mut}$
and $t_\text{fix}$ becomes blurred, which signals the onset of
clonal interference.
}
\end{figure}

On the other hand, if $t_\text{mut}$ and $t_\text{fix}$ are of the
same order, then mutations can occur during the fixation process of 
previous mutations~\cite{MS1971,W2004} and the distinction between $t_\text{mut}$
and $t_\text{fix}$ becomes unclear. In Fig.~\ref{Fig:CI}, we present 
an example showing how the population dynamics changes when the criterion
(\ref{onset}) is violated. Following \cite{GL1998} we refer to the interaction among
beneficial mutant clones in this regime as \textit{clonal interference}. 

In the remaining parts 
of this section we present the main analytic approaches that have been developed to compute
the speed of evolution in the clonal interference regime. We begin by considering the case
where all mutations have the same selection coefficient 
(\textit{model I}) and then
treat the case of a continuous distribution of selection coefficients\footnote{Note that in part of the literature
\cite{DF2007,Brunet2008,Rouzine2008} the term ``clonal interference''
is restricted to model II.} (\textit{model II}).

\subsection{\label{Sec:single}Model I: Single selection coefficient of beneficial mutations}

\subsubsection{The Crow-Kimura-Felsenstein approach}
\label{CKF}
The first attempt to compute the speed of evolution in the presence of clonal
interference is due to Crow and Kimura \cite{CK1965}.
We present their calculation 
in the form given by Felsenstein \cite{Felsenstein1974}, which takes 
into account that only established mutations contribute to the
adaptation process. Such mutations (with selection coefficient $s_b$) appear in the population at rate
$\pi(s_b) N U$. Assuming that a mutation was established at time $t=0$, 
we now ask for the waiting time $\tau$ until a second mutation is
established \textit{in the offspring of the first}. 
We take $s_b$ to be small, such that $\pi(s_b) \approx 2s_b$ and $(1+s_b)^t \approx e^{s_bt}$.
Then, according to (\ref{abclone}), 
the frequency $a_t$ of the mutant starting at $a_0 = 1/(2s_bN)$ is
\begin{equation}
\label{clone}
a_t = \frac{1}{1 + (2 s_bN - 1)e^{-s_bt}}.
\end{equation}
The number of mutants at time $t$ is $N a_t$, and each mutant
generates an established second mutant with probability $2 s_b U$ per
generation. We therefore need to compute the 
accumulated number of mutants $N_\text{acc}$ that have existed up to time $t$, where each
individual is weighted by the number of generations during which
it has existed. Approximating the sum over generations by an
integral, this is given by 
\begin{equation}
\label{integrate}
N_\text{acc}(t) \simeq N \int_0^t dt' \; a_{t'} =
\frac{N}{s_b} \ln \left[ \frac{e^{s_bt}}{2 Ns_b} + 1 - \frac{1}{2 Ns_b}
\right]
\approx \frac{N}{s_b} \ln \left[ \frac{e^{s_bt}}{2 Ns_b} + 1 \right]
\end{equation}
for $N s \gg 1$.
The waiting time $\tau$ is then determined from the condition 
$2 s_b U N_\mathrm{acc}(\tau) = 1$, which yields the speed 
\begin{equation}
\label{Felsenstein}
v_N^\text{CKF} = \frac{s_b}{\tau} = \frac{s_b^2}{\ln[(2Ns_b)(e^{1/2UN}-1)]}.
\end{equation}
For small $N$ ($U N \ll 1$) this reduces to the expression
(\ref{periodic}) valid in the periodic selection regime,
while for large $N$ a finite speed limit
$v_\infty = s_b^2/\ln(s_b/U)$ is reached.

In writing the relation (\ref{Felsenstein}) it is implicitly assumed that the
situation at the apperance of the second mutation is identical to that
at the appearance of the first, which is not true: the second mutation competes
against a background consisting of a mixture of mutant and wild type with 
mean fitness (relative to the wild type fitness of unity)
$\bar w = (1+s_b) a_t + (1-a_t) = 1+a_t s_b < 1 + s_b$. The selective advantage of the second 
mutant compared to the background population is therefore larger than $s_b$, and it will
grow faster than the first mutant population. For this reason the expression (\ref{Felsenstein})
is a lower bound on the actual speed. To improve on this bound we need to take into account
the coexistence of several mutant clones in the population, which will be the subject of the 
next subsection.

\subsubsection{The traveling wave approach}
As the discussion in Sec.~\ref{Sec:GLinf} shows, the deterministic evolution of
an infinite population is well described as a travelling wave of approximately
Gaussian shape. In order to extend this approach to large but finite populations,
the deterministic dynamics of the bulk of the wave is combined with a stochastic
description of the appearance of new mutants at the high-fitness edge of the 
frequency distribution. This idea was first proposed by Rouzine, Wakeley and Coffine
\cite{Rouzine2003} and has since been further elaborated \cite{DF2007,Rouzine2008,Brunet2008}.
In this section we follow the particularly simple and transparent derivation presented
in \cite{Beeren2007}.

As in Sec.~\ref{Sec:GLinf}, we denote by $f_t(n)$ the frequency of individuals with 
$n$ mutations, and assume for this distribution the Gaussian form
\begin{equation}
f_t(n) \approx \frac{1}{\sqrt{2 \pi \sigma^2}} 
\exp\left (-\frac{(n-v_N t/s_b)^2}{2 \sigma^2} \right ).
\label{Eq:fin_gauss}
\end{equation}
Here we have used that the mean number of mutations acquired up to time $t$ is $\bar n = v_N t/s_b$.
The speed $v_N$ and the variance $\sigma^2$ of the travelling wave are related by Fisher's fundamental
theorem or, more generally, by the Guess relation \eqref{eq:guessrelation}.
Neglecting the direct mutation contribution $U \ln (1 + s_b)$ because of $U \ll 1$ and evaluating
the selection term using the approximation $\chi_i \approx 1 + s_b(n_i - \bar n)$ (where $n_i$ is 
the number of mutations acuired by individual $i$) we see that
\begin{equation}
v_N \approx s_b^2 \sigma^2,
\label{Eq:v_sigma}
\end{equation}
which is also true for the infinite population case when $U \ll 1$ (compare to 
(\ref{Eq:var_ben})). 

It is clear that at any finite time $t$, there is a maximal number of mutations 
$n_\text{max}(t)$ such that
$f_t(n) = 0$ for $n > n_\text{max}(t)$. Let
\begin{equation}
L(t) \equiv n_\text{max}(t) - \frac{1}{s_b} \ln \bar w(t)
\end{equation}
denote the \textit{lead} of this class of fittest individuals relative to the
mean population fitness \cite{DF2007}.
Let $t_n$ be the generation when $n_\text{max} = n$ for the first time.
We assume that $\langle L(t_n) \rangle \rightarrow L_0$ as $t\rightarrow 
\infty$ and $\langle t_{n+1} - t_n \rangle \rightarrow \tau$, with constant $L_0$ and $\tau$, 
which reflects the existence of a stationary travelling wave with
speed 
\begin{equation}
\label{speed}
v_N = s_b/\tau,
\end{equation}
compare to (\ref{Felsenstein}). 
For times $t_n < t < t_{n+1}$, $L(t)$ then behaves as
$L_0 - \ln \bar w/s_b \simeq L_0 - v_N t/s_b$.
We further assume that, for very large $N$, the lead satisfies $L(t) s_b \gg 1$,
which implies that the loss by genetic drift of new mutants arising from the most fit class
can be neglected. Analogous to Sect.\ref{CKF}, we can now compute the
accumulated number of mutants in the most fit class  
that have existed during the time $t_n < t < t_{n+1}$ according to 
\begin{equation}
N_\text{acc}(\tau) = \sum_{t=1}^\tau \exp\left ( s \sum_{u=1}^t L(u) \right ) 
= \sum_{t=1}^\tau \e^{( L_0 s_b t - \frac{1}{2}v_N t^2)} \approx
\frac{\e^{L_0 s_b \tau} - 1}{1-\e^{-L_0 s_b}} \approx \e^{L_0 s_b \tau} ,
\end{equation}
which is a good approximation if $v_N \tau = s_b \ll L_0 s_b$ or $L_0 \gg 1$.

The mean waiting time until the appearance of a mutant with
$n_\text{max}+1$ mutations 
is the solution of the equation $N_\text{acc}(\tau) U = 1$, which yields
\begin{equation}
\tau \approx \frac{1}{L_0s_b}  \ln \left ( \frac{1}{ U}\right ).
\label{Eq:tau_q}
\end{equation}
Finally, we close the system of relations by noting that, as long as
$N$ is not extremely large\footnote{More precisely, $N \ll N_c$.}, the
new fittest class will most likely appear as a single individual,
which implies that
\begin{equation}
\frac{1}{\sqrt{2 \pi \sigma^2}} \e^{-\frac{L_0^2}{2 \sigma^2}} = \frac{1}{N}
\rightarrow L_0 s_b  = \left ( 2 v_N \ln \frac{N s_b}{\sqrt{ 2 \pi v_N}} \right )^{1/2}.
\label{Eq:sigma_N}
\end{equation}
From Eqs.~\eqref{Eq:v_sigma}, \eqref{Eq:tau_q}, and \eqref{Eq:sigma_N},
$v_N$ becomes the solution of the equation
\begin{equation}
v_N \simeq \frac{ 2 s_b^2 \ln \left ( \frac{Ns_b}{\sqrt{2 \pi v_N}} \right
  ) }{(\ln U )^2}, 
\label{Eq:vsol1}
\end{equation}
which leads to the final result
\begin{equation}
\label{Eq:vsol}
v_N^\text{Gauss} = \frac{2 s_b^2 \ln (N)}{(\ln U)^2}
\end{equation}
for very large $N$.
The logarithmic growth of the speed with $N$ must saturate when the 
infinite population limit $v_N = s_b$ is approached. According to 
(\ref{Eq:vsol}) this happens when $N \sim N_c \sim e^{(\ln U)^2/2 s_b}$, in
agreement with the estimate \eqref{Nc}. For population sizes exceeding
$N_c$ the relation (\ref{Eq:sigma_N}) is no longer valid, because  
the initial frequency of the fittest genotype at $t_n$ can be much larger
than $1/N$ once $N \gg N_c$. 
The existence of an absolute speed limit $v_N = s_b$ is evident from
(\ref{speed}), because $\tau$ cannot be smaller than one generation
time in the discrete time model.
For models with overlapping generations such a restriction does not
exist, because a larger number of offspring can be generated 
within much less than an average generation time, and the speed
increases proportional to $\ln N$ for arbitrary $N$.

In this context, it is instructive to compare the discrete and the
continuous time models in different population size regimes.
When the population size is small ($N U \ll 1$), there is 
a slight difference between these two models. For example, the fixation
probability for small $s$ is $C s$ with a model dependent constant $C$ (compare
(\ref{Moranfix}) and (\ref{Kfix})). Once the
population becomes large enough so that the loss of the fittest
type by genetic drift can be neglected, there is no difference between
the continuous and discrete time models. However, for very large $N \ge
N_c$, there is a large difference 
due to the restriction $\tau \geq 1$ in the discrete time model.

\subsubsection{Comparison to simulations}

\begin{figure}
\centerline{\includegraphics[width=0.7\textwidth]{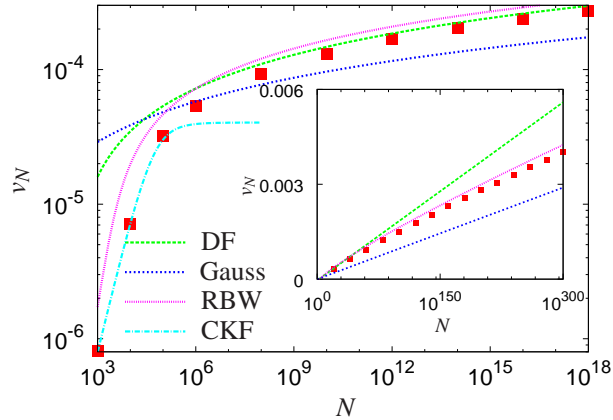}}
\caption{\label{Fig:comall} Comparison of the theoretical expressions
Eqs.~\eqref{Eq:DF}, \eqref{Eq:vsol}, \eqref{Eq:RBW}, and \eqref{Felsenstein}  
with simulations using $s_b=0.02$ and $U=10^{-6}$. In the inset the comparison
is extended up to $N = 10^{300}$ except for Eq.~\eqref{Felsenstein}, which
predicts a speed limit. The algorithm used to obtain these data
is described in the Appendix.
}
\end{figure}

The above derivation of the speed of evolution involves a number of
rough, uncontrolled approximations, such that the result
(\ref{Eq:vsol}) can hardly be expected to be quantitatively accurate.
A much more careful analysis along similar lines was presented by 
Rouzine, Brunet and Wilke (RBW) \cite{Rouzine2008}, 
who find the implicit expression\footnote{The speed $V$ in Ref.~\cite{Rouzine2008} is 
$v_N/s_b$. To conform to our notation,
we slightly modified Eq.~(52) in Ref.~\cite{Rouzine2008}.} 
\begin{equation}
\ln N \approx \frac{v_N^\text{RBW}}{2 s_b^2}
\left ( \ln^2 \frac{v_N^\text{RBW}}{\e U s_b} + 1 \right ) 
-\ln \sqrt{\frac{s_b^3 U}{v_N^\text{RBW} \ln (v_N^\text{RBW}/(U s_b))}}.
\label{Eq:RBW}
\end{equation}
A related approach,
which however does not explicitly use the Gaussian shape of the
deterministic part of the travelling wave, was presented by Desai 
and Fisher \cite{DF2007}, who find\footnote{A detailed analysis of
this approach can be found in \cite{Brunet2008}.}
\begin{equation}
v_N^\text{DF} \approx \frac{ 2 s_b^2 \ln N}{ \ln^2 (U/s_b)}.
\label{Eq:DF}
\end{equation}
In Fig.~\ref{Fig:comall}, we compare the different theories with 
simulation results for the WF model. 
For moderately large population size, Eqs.(\ref{Eq:RBW}) and 
(\ref{Eq:DF}) are of comparable quality, but for extremely large population,
as shown in the inset of Fig.~\ref{Fig:comall}, the predictive
power of Eq.~\eqref{Eq:RBW} is superior to the other approaches.
 
In the asymptotic regime, Eq.~\eqref{Eq:RBW} predicts that
$v_N \sim \ln N / \ln^2 \ln N$, but Eqs.~\eqref{Eq:vsol} and 
\eqref{Eq:DF} predict
$v_N \sim \ln N$. Rigorous work~\cite{Yu2007,Yu2008}
has established that the speed in the asymptotic regime is not smaller than
${\cal{O}}({\ln^{1-\delta} N})$ for any positive $\delta$, which does not 
exclude the
possibility of a multiplicative $\ln^2 \ln N$-correction. 
Even with the dedicated algorithm used to generate the data in the 
inset of Fig.~\ref{Fig:comall}, it seems hardly
possible to settle this issue using numerical simulations.

\subsection{Model II: Continuous distribution of selection coefficients}
In Sect.~\ref{Sec:single}, we reviewed theories aimed at calculating 
the speed of evolution when the selection coefficient takes a single value (model I). 
In this subsection, we will allow the selection coefficient 
to take a continuous range of values drawn from a distribution like
Eq.~\eqref{eq:def:gbeta} (model II). Unlike model I, two mutants
arising from the same progenitor now have different selection coefficients
and selection is operative between these two mutations. In contrast,
in model I the competition between two mutants derived from a single
progenitor is purely stochastic, and selection operates only between
clones that have accumulated a different \textit{number} of mutations.

The qualitative picture of a wave of fixed shape traveling along the
fitness axis that we developed for model I is expected to apply 
to model II as well, but it is more difficult to quantify,
because continuous fitness cannot be reduced to the discrete 
number of acquired mutations. Two approaches have so far been proposed
to deal with this problem. The first is related to an ``equivalence
principle'' discovered in microbial evolution
experiments \cite{Hegreness2006}, which suggests that a given
distribution of selection coefficients can be represented by an
effective single selection coefficient along with a suitably rescaled
effective mutation rate. A heuristic scheme to implement this idea was
given in \cite{DF2007} and tested against numerical simulations
in \cite{Fogle2008}. As one might expect, the representation 
by a single ``predominant'' selection coefficient is 
quantitatively accurate only if the distribution is very narrow, such as 
$g^{(\beta)}$ with $\beta = 10$, and it fails completely when $\beta
\leq 1$ \cite{Fogle2008}.

The second approach, first proposed by Gerrish and Lenski (GL)
\cite{GL1998}, attempts to extend the periodic selection picture into
the clonal interference regime by focusing on mutations of
exceptionally large effect. Clonal interference is seen as a filter
that eliminates mutations whose effect is small enough to be superseded by 
a mutation of larger effect arising later in the process. Once the
size of selection coefficients of mutations that survive the competition by
other clones has been identified, along with the rate at which such
mutations appear, the speed of evolution is obtained from a simple
relation similar to Eq.~(\ref{periodic}) used in the periodic selection regime. 

In the following we outline the GL approach, derive its asymptotic
predictions, and compare it to simulation results.

\subsubsection{Gerrish-Lenski Theory}

The GL-theory is based on two assumptions \cite{GL1998,PK2007}. 
First, the type of any individual at any time is either the wild type or
a mutant derived directly from the wild type. 
The contributions from \textit{multiple} mutations arising 
from an extant mutant are neglected.
Since the fixation of a mutation under this assumption becomes
a renewal process~\cite{G2001}, we will refer to this assumption as
the renewal assumption.
Second, the loss of a beneficial mutation by stochastic sampling error
when rare (genetic drift)  is 
determined solely by its selection coefficient compared to the wild type.
Other beneficial mutations do not play any role in determining the
fate of the mutation at early times. We will refer to this assumption as the assumption
of establishment. 

\begin{figure}
\includegraphics[width=\textwidth]{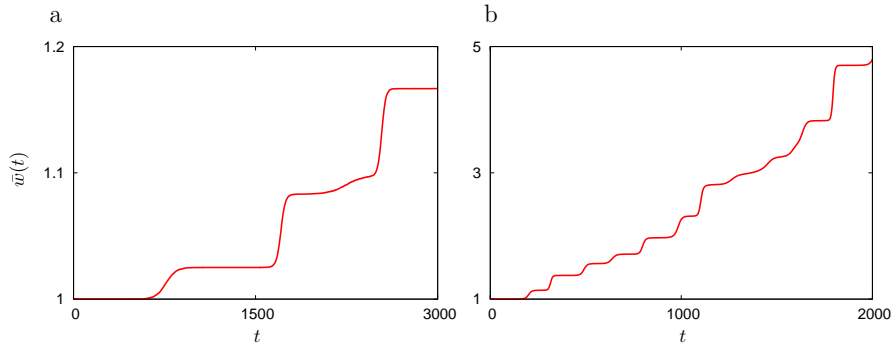}
\caption{\label{Fig:glassume} Plots of mean fitness corresponding 
to the two panels on the right hand side of Fig.~\ref{Fig:CI}.
The population sizes are (a) $10^5$ and (b) $10^7$. 
Although assumptions of the GL approach are not strictly applicable,
one observes regions where the mean fitness behaves in a step-like fashion.
}
\end{figure}

The picture underlying these two assumptions is that the adaptive process
can still be decomposed into separate selective sweeps in which a
mutation grows in a fixed background and eventually takes over the
population (compare to Fig.\ref{Fig:periodic}). A signature of this kind of
dynamics is a step-like increase of the mean fitness. 
As can be seen in Fig.~\ref{Fig:CI}, this step-like behavior is 
pronounced for small populations in the periodic selection regime. 
However, as the population size increases, the mean fitness
becomes more and more smooth, see Fig.~\ref{Fig:glassume}, although
distinct steps still occur when a mutation of exceptionally large
strength appears.
Thus, the GL approach is expected to be useful in a 
restricted range of population sizes, which goes slightly
beyond the periodic selection regime. It is similar in spirit to the 
Crow-Kimura-Felsenstein approach reviewed in Sect.\ref{CKF}, which also successfully
captures the slowing down of adaptation near the onset of clonal interference
but fails for larger $N$ (compare to Fig.\ref{Fig:comall}). 

To formulate the GL-theory quantitatively, we make use of two functions
introduced previously: the probability distribution $g(s)$ 
of selection coefficients (like $\gbeta$ in Eq.~\eqref{eq:def:gbeta}), and 
the probability $\pi(s)$ for the fixation (or, equivalently, the
establishment) of a mutation of strength $s$.
By the assumption of establishment, the distribution of the mutations that 
can spread in the population after the initial fluctuations and are really 
competing is then given by\footnote{Without the
assumption of the establishment, the survival probability of a mutation 
should also depend on the population structure at the time
when this mutation arises.} $\pi(s) g(s)$.

\begin{figure}
\includegraphics[width=\textwidth]{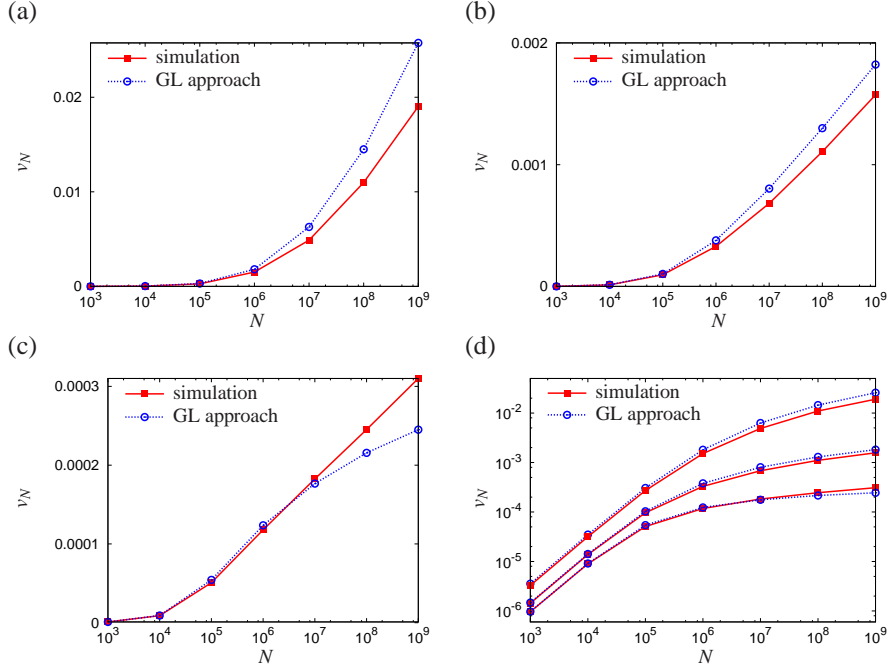}
\caption{\label{Fig:GL} Comparison of the GL theory with the simulation 
results of the WF model for $U=10^{-6}$ and mean selection coefficient
$\Gamma(1+1/\beta) s_b = 0.02$ for
(a) $\beta=\frac{1}{2}$, (b) $\beta=1$, and (c) $\beta = 2$. Panel (d) shows the 
data from (a)-(c) in double logarithmic scales.}
\end{figure}

For a mutation with selection coefficient $s$ to be fixed, 
it is necessary that no fitter mutation is established during the
time required for the first mutation to fix. 
The expected number of established fitter mutations that appear during
this time is 
\begin{equation}
 \lambda(s) = \left(NU t_\text{fix}(s)/2\right ) \int_s^{\infty} du \pi(u)g(u)
\label{Eq:meanfix}
\end{equation}
where\footnote{
Note that in previous work on the GL approach the expression $t_\text{fix} = 2 \ln N /s$
was used irrespective of the size of $s$~\cite{GL1998,W2004,PK2007}. We will get back
to the consequences of this replacement in Sect.~\ref{Sec:GLasym}.} $t_\text{fix}(s)$
is given in Eq.~\eqref{Eq:Fix}, the factor 1/2 
comes from the renewal assumption [see also discussion below Eq.~(\ref{Eq:AB})], and the integral gives the probability
that the selection coefficient of an established mutation is larger than $s$.
Note that the renewal assumption prohibits a secondary mutation with selection
coefficient $s''$ arising in the offspring of a primary mutation $s'$ with 
$s' < s$ but $s' + s'' > s$, which would 
make Eq.~\eqref{Eq:meanfix} much more complicated.
Hence, within the GL approximation the probability 
of not encountering any fitter mutation during fixation is $\exp(-\lambda(s))$
and, accordingly,
the fixation probability of a mutation with selection coefficient $s$ becomes
\begin{equation}
\label{Eq:Pfix}
P_\text{fix}(s) = 
\pi(s) g(s) \exp
\left ( - \frac{N U \ln N}{\ln (1+s)} \int_s^\infty \pi(u) g(u) du \right ).
\end{equation}
In words, for a mutation with selection coefficient $s$ to be fixed,
it must first survive genetic drift (with probability $\pi(s) g(s)$), 
then should outcompete all other mutations 
(with probability $\exp(-\lambda(s))$).
Thus, the substitution rate (the number of fixed mutations
per generation) is
\begin{equation}
\label{Eq:keff}
k_\text{eff} = N U \int_{s=0}^\infty P_\text{fix}(s) ds.
\end{equation}
To calculate the speed $v_N$, we need the mean selection coefficient
of fixed mutations which is readily obtained as
\begin{equation}
\label{Eq:seff}
s_\text{eff} = \frac{\int s P_\text{fix}(s) ds}{\int P_\text{fix}(s) ds}.
\end{equation}
Along with $k_\text{eff}$ this determines the speed according
to\footnote{The reader may wonder why $\ln(1+s_\text{eff})$ on the
  right hand side of (\ref{eq:gerrishlenskivelocity}) is not replaced
  by the average of $\ln(1+s)$ with respect to $P_\text{fix}(s)$. For
  small $s_\text{eff}$ the difference between the two is obviously
  neglible, but the same is true when $s_\text{eff} \gg 1$, because
  then $P_\text{fix}$ becomes very narrow due to clonal inteference. For a numerical
test of (\ref{eq:gerrishlenskivelocity}) see \cite{PK2007}.}~\cite{W2004}
\begin{equation}
\label{eq:gerrishlenskivelocity}
v_N^\text{GL} = k_\text{eff} \ln (1 +s_\text{eff} ).
\end{equation}
In Fig.~\ref{Fig:GL}, we compare Eq.~\eqref{eq:gerrishlenskivelocity} to simulations 
using $\gbeta$ with three different values of $\beta$.
The integrals in (\ref{Eq:Pfix}), (\ref{Eq:keff}) and (\ref{Eq:seff}) were evaluated numerically. 
We see that the GL approach is remarkably accurate also beyond the periodic selection regime,
as becomes evident by comparing the double-logarithmic graph in Fig.\ref{Fig:GL}(d) to the 
corresponding Figure \ref{Fig:comall} for model I. 
However the deviations grow as $N$ increases, in particular for $\beta = 2$, where
the GL-prediction shows a negative curvature in $\ln N$ which is not present in the simulation
data. We will return to this point at the end of the next subsection. As the numerical scheme
employed for model I relies on the discreteness of the fitness space (see Appendix), we have
no information about the behavior of $v_N$ for very large $N$. Since we have shown 
in Sect.\ref{Continuum} that the speed of evolution is infinite in the infinite population
model, we merely know that (in contrast to model I) $\lim_{N \to \infty} v_N = \infty$. 

\subsubsection{\label{Sec:GLasym}Asymptotic Behavior of the GL theory}
Although we cannot analytically evaluate the expression for $v_N$ predicted
by the GL approach, an accurate asymptotic approximation 
can be derived, which is the topic of this section.
Throughout the distribution $g^{(\beta)}$ of selection coefficients is used.
The calculation follows the idea presented (for $\beta = 1$) in 
Ref.~\cite{PK2007}; see also Ref.~\cite{W2004}. 
The only difference is that $t_\text{fix} = 2 \ln N /\ln(1+s)$ is used rather
than $2 \ln N /s$, which will turn out to affect the conclusion significantly.

The integrations involved in the GL theory take the form
\begin{equation}
I[A;n] = \int_0^\infty ds s^n f(s) \exp(-A h(s)),
\end{equation}
where $f(s) = \pi(s) \gbeta (s)$, $h(s) = \int_s^\infty f(u) du /\ln(1+s)$,
and $A =  N U\ln N$. Note that $h(s)$ is a decreasing function with the range
$[0,\infty]$.
By the change of variable $y = h(s)$, the above integral becomes
\begin{equation}
I[A;n] = \int_0^\infty \Psi(y) \e^{-A y}dy ,
\label{Eq:IAn}
\end{equation}
where
\begin{equation}
\Psi(h(s)) =  
s^{n} \frac{f(s)}{|h'(s)|} 
=  s^{n} \ln (1+s) + \frac{s^n}{1+s} \left ( \frac{d}{ds} \ln h(s) \right)^{-1}.
\end{equation}
To arrive at Eq.~\eqref{Eq:IAn}, we 
have used $f(s) = -(d/ds)(\ln(1+s) h(s))$ and the fact that $h(s)$ is a
(monotonic) decreasing function.
As $A\rightarrow \infty$, $I[A;n]$ is dominated by the contribution
around $y=0$, or equivalently around $s=\infty$. 
When $s$ is very large, we can approximate $\pi(s) \approx 1$ and hence
$h(s) \sim \exp(-(s/s_b)^\beta)/\ln s$.
Hence for large $s$, we can approximate $s = s_b (- \ln y)^{1/\beta}$ 
($y \ll 1$), and
\begin{equation}
\Psi(y) \approx s^{n}\ln s - s^{n-1} \frac{s_b}{\beta} \left ( \frac{s_b}{s}
\right)^{\beta-1}
\approx s_b^{n} (-\ln y)^{n/\beta} \ln (\ln(1/y))/\beta,
\end{equation}
where we have kept only the leading order term.
Hence
\begin{eqnarray}
I[A;n] \approx \frac{s_b^{n}}{\beta} \int_0^\infty (-\ln y)^{n/\beta} \ln \ln(1/y) \e^{-A y} dy
\approx \frac{s_b^{n}}{A\beta} (\ln A)^{n/\beta}\ln \ln A.
\end{eqnarray}
The substitution rate is then
\begin{equation}
\label{Eq:GLk}
k_\text{eff} = N U I[N U \ln N;0] = \frac{\ln \ln (N U \ln N)}{\beta \ln N},
\end{equation}
which, as $N\rightarrow \infty$, 
approaches 0  for any $\beta$.
The asymptotic behavior of the speed is
\begin{equation}
v_N^\text{GL} =k_\text{eff} \ln \left (1 + \frac{I[A;1]}{I[A,0]}\right )
\sim \frac{\left ( \ln \ln N \right )^{2} }{\beta^2 \ln N},
\label{Eq:GLV}
\end{equation}
which also approaches 0 as $N\rightarrow \infty$.

\begin{figure}
\centerline{\includegraphics[width=0.6\textwidth]{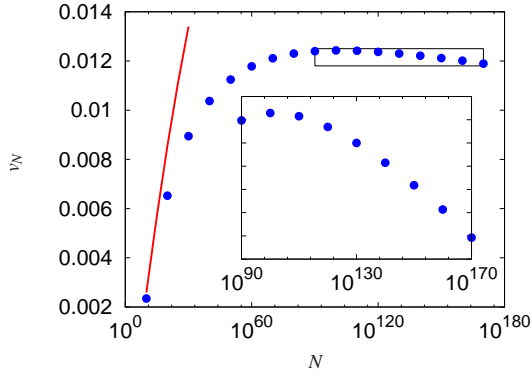}}
\caption{\label{Fig:GLcom} Comparison of the GL theory using $t_\text{fix} = 2 \ln N / \ln(1+s)$ (symbols) with that using $t_\text{fix} = 2 \ln N /s$ 
(line) for $U=10^{-6}$, $s_b = 0.02$, and $\beta = 1$. Inset: Close up
of the boxed area. As anticipated by the asymptotic analysis, the speed
decays, though slowly, with $N$. 
}
\end{figure}
This asymptotic behavior is easily understandable from an
extremal statistics argument.
The maximal mutation coefficient $s_\text{max}$ observed over ${\cal M}$ mutation events is given approximatively as a solution of
\begin{equation}
 \text{Prob}\left\{ s> s_\text{max}\right\} = \int_{s_\text{max}}^\infty g(u)du
=\exp\left (-(s_\text{max}/s_b)^\beta\right ) \simeq \frac{1}{{\cal M}}.
\end{equation}
Following the GL hypothesis, the selection coefficient that gets 
fixed has to be the maximum of all selection coefficients that appear 
within its own fixation 
time, i.e. one has to consider a typical number
\begin{equation}
\label{Eq.:M} 
{\cal M} \sim NU t_\text{fix} \sim NU \ln N / \ln(1+s_\text{max})
\end{equation} of mutations. 
Thus, the leading behavior of $s_\text{max}$ becomes
$s_\text{max} \sim s_b \ln^{1/\beta} A \sim s_b \ln^{1/\beta} N$, the effective substitution rate is given by (up to leading order)
\begin{equation}
k_\text{eff}\approx 1/t_\text{fix}(s_\text{max}) \sim \frac{\ln \ln(N)}{\beta \ln N}
\end{equation}
as in Eq.~\eqref{Eq:GLk}, and the velocity is the same as in Eq.~\eqref{Eq:GLV}.

The asymptotic behavior obtained in this section is 
completely different from previous reports\footnote{Note that in the original 
paper of Gerrish and Lenski \cite{GL1998}, it was erroneously concluded that 
$v_N$ approaches a finite ``speed limit'' for $N \to \infty$.}~\cite{GL1998,W2004,PK2007}.
The reason is clearly the factor $\ln (1+s)$ in the
denominator of $t_\text{fix}(s)$, which is very different from $s$  when 
$s_\text{eff} \gg 1$.
However, this effect is only relevant when $N$ is extremely large. 
As Fig.~\ref{Fig:GLcom} shows, the true asymptotic behavior (\ref{Eq:GLV}) is 
approached only when $N \gg 10^{100}$ for $U=10^{-6}$ and $s_b = 0.02$ with $\beta = 1$, and
the difference between using $\ln(1+s)$ and $s$ in $t_\text{fix}(s)$ is small
when $N \le 10^{20}$. So for realistic values of $N$,
replacing $\ln (1+s)$ by $s$ 
can provide a good approximation for the speed. 

In fact, if the mutant fitness is derived
from the parental fitness by multiplication with 
$\e^s$ rather than with $1+s$, which would correspond to a continuous time picture, the
fixation time is $2 \ln N/s$ for all $s$. The speed is then given by 
the expression $v_N = k_\text{eff} s_\text{eff}$ used in
Ref.~\cite{GL1998}, rather than by Eq.~\eqref{eq:gerrishlenskivelocity}.
The leading asymptotic behavior of GL theory within this scheme 
can be obtained along the lines of \cite{PK2007} or, 
more directly, by adapting the extremal statistics argument given above.
Since the leading behavior
of $s_\text{max}$ is the same as before, the asymptotic behavior becomes
\begin{equation}
\label{Eq:GLs}
k_\text{eff} \sim s_\text{max}/\ln N \sim s_b \ln^{1/\beta-1} N,
\quad
v_N^\text{GL} \sim s_b^2 \ln^{2/\beta - 1} N.
\end{equation}
Thus the graph of $v_N$ versus $\ln N$ within GL theory is positively curved when
$\beta < 1$ and negatively curved when $\beta > 1$, as is visible
in Figs.\ref{Fig:GL} (a)-(c). The simulation results for $\beta = 2$ do however not share
this feature, and lie distinctly above the GL prediction for large $N$.
In the next subsection we elaborate on this observation.

\subsubsection{Importance of multiple mutations}

It is instructive to compare (\ref{Eq:GLs}) to the result
$v_N \sim s_b^2 \ln N$ obtained for model I in Sect.\ref{Sec:single}. Evidently,
the speed in model I should be minimal among all distributions $g(s)$ with the same
mean selection coefficient, which implies that $v_N$ should increase at least 
as fast as $\ln N$ also for model II\footnote{For the purpose of this discussion we
ignore the saturation of the speed that occurs at extremely large $N$ in model I.}.
However, according to (\ref{Eq:GLs}) $v_N$ grows more slowly than $\ln N$ when $\beta > 1$,
and even decreases with $N$ when $\beta > 2$. Moreover, the rate of substitution 
decreases with increasing $\ln N$ for $\beta > 1$, although we know that $k \to 1$ in the 
infinite population limit. 

This is not really surprising, as the GL approach
takes into account only the mutations of largest effect, ignoring the cumulative effect
of multiple mutations of average effect which drive the dynamics in model I.
On the basis of (\ref{Eq:GLs}), one might speculate that the evolutionary process
is dominated by large, extremal selection coefficients when $\beta < 1$,
and by multiple mutations of typical effect when $\beta > 1$. This could also account for
the breakdown of the ``predominant mutation'' approach for $\beta < 1$ \cite{DF2007,Fogle2008}.
Interestingly, the exponential distribution of selection coefficients, which is most widely
used in this context \cite{GL1998,O2003,W2004,PK2007}, would then turn out to represent the marginal case separating
the two regimes. 

\begin{figure}
\centerline{\includegraphics[width=0.7\textwidth]{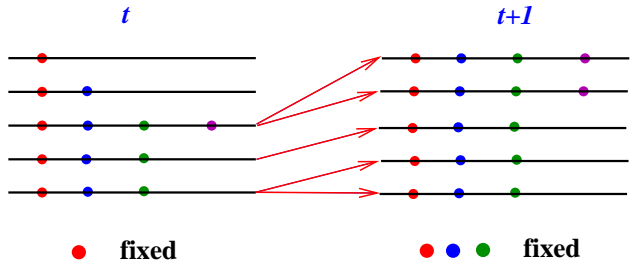}}
\caption{\label{Fig:multi} Fixation of multiple mutations in a population of size $N=5$. At time $t$, four
types are present, and only the red mutation is fixed (= shared by all individuals). In the next
generation, the individuals with one and two mutations leave no offspring, and consequently the 
blue and the green mutation go to fixation simultaneously.
}
\end{figure}

\begin{figure}
\includegraphics[width=0.49\textwidth]{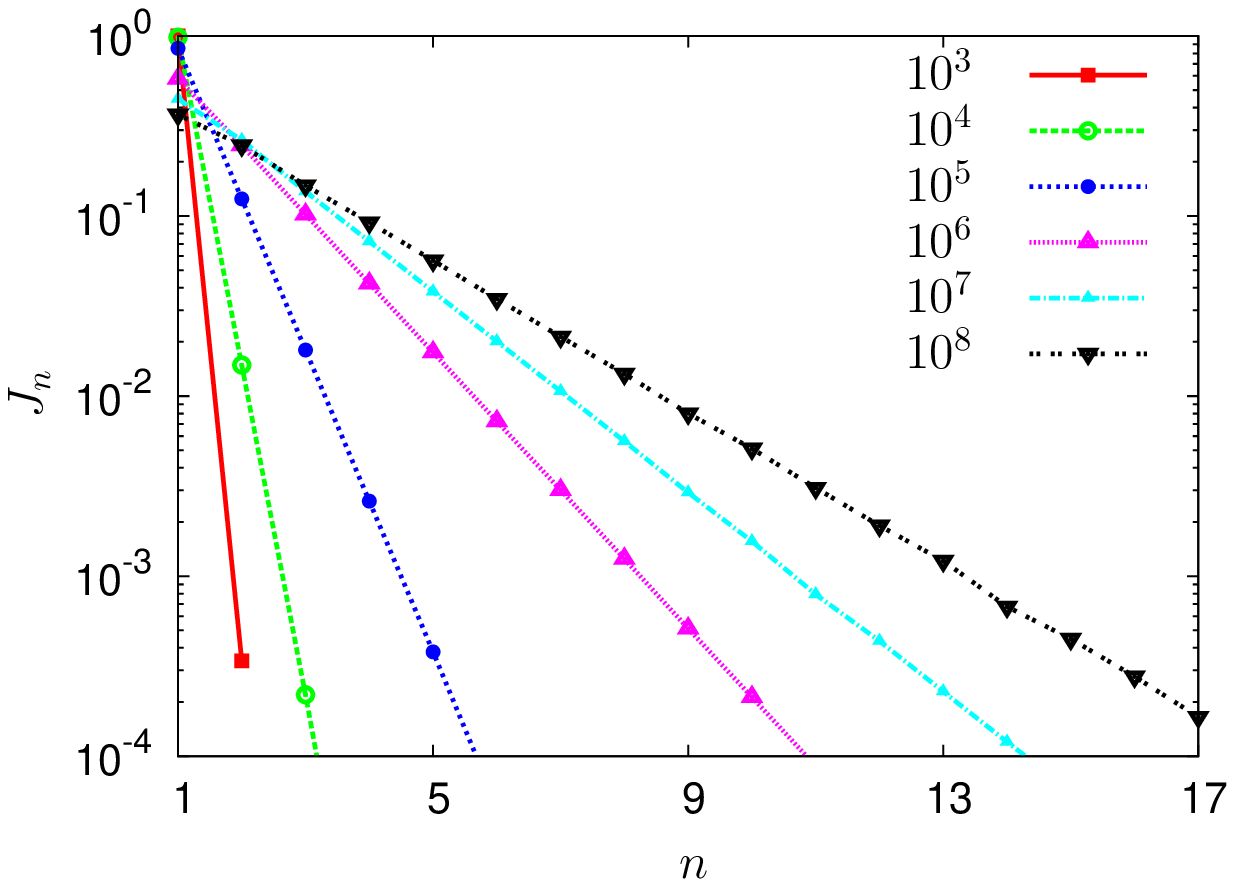}
\includegraphics[width=0.49\textwidth]{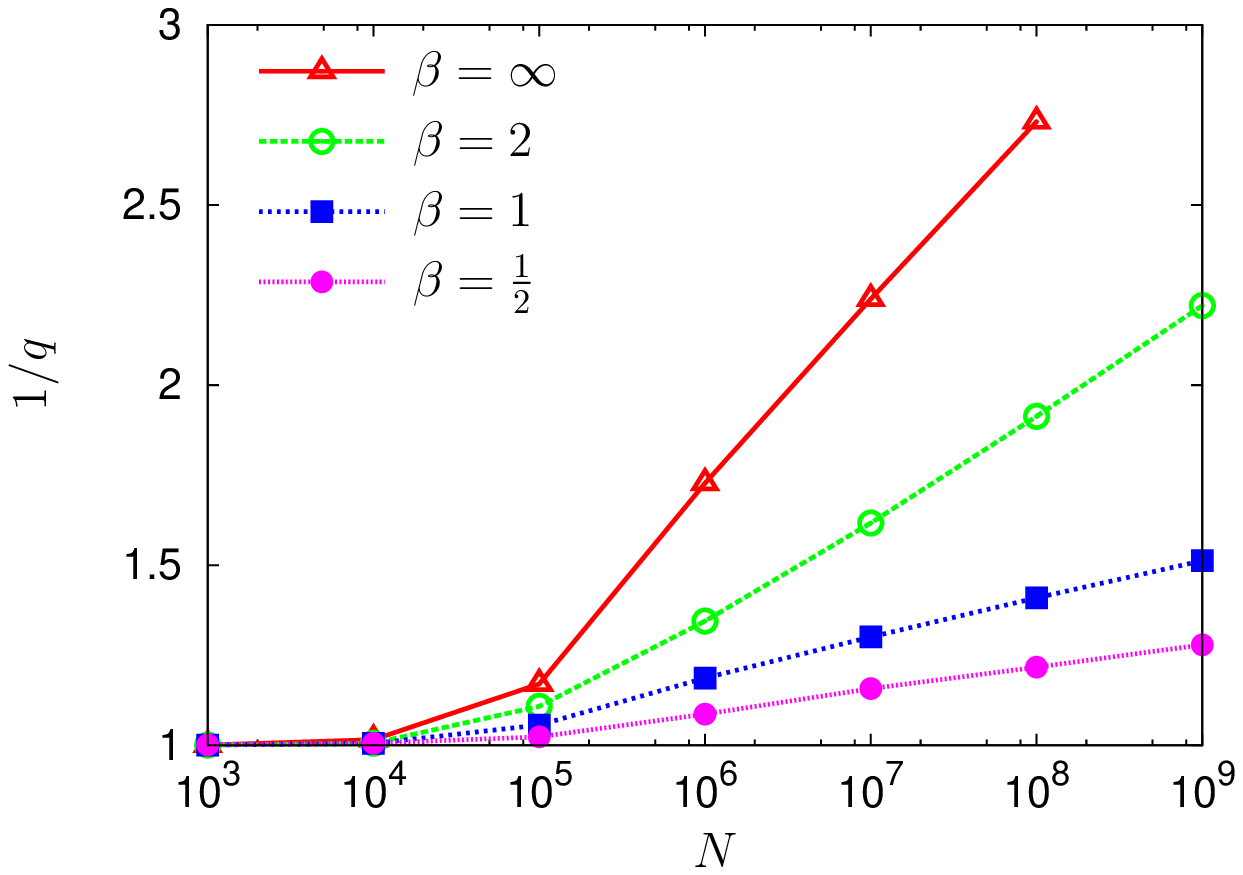}
\caption{\label{Fig:q} Left : 
Distribution of the number of fixed mutations
per fixation events for $\beta =\infty$ with $s_b = 0.02$ and
$U = 10^{-6}$ in semi-logarithmic scales.
From left to right, the population sizes are $10^3$, $10^4$,
$10^5$, $10^6$, $10^7$, and $10^8$. Clean geometric distributions
are observed.
Right : Mean number of fixed mutations per fixation event ($1/q$) as
a function of population size.
}
\end{figure}

A quantitative measure of the importance of multiple mutations in the evolutionary dynamics
can be obtained by asking how many mutations typically go to fixation in a single fixation
event. The way in which the fixation of different mutations can become linked is illustrated
in Fig.\ref{Fig:multi}. It was observed numerically in \cite{PK2007} (for model II with 
$\beta = 1$) that the probability 
$J_n$ of $n$ mutations fixing in a single event is well described by a geometric distribution,
\begin{equation}
\label{Eq:Jn}
J_n = (1-q)^{n-1} q.
\end{equation}
The left panel of Fig.\ref{Fig:q} shows that the same relationship holds for model I. 
The parameter $1/q$ is the mean number of simultaneously fixed mutations, and it
increases with $N$ in a logarithmic fashion (right panel of Fig.\ref{Fig:q}).
As expected, mutiple mutations are more prevalent for larger $\beta$, but
there does not seem to be any qualitative difference in the behaviors for $\beta < 1$
and $\beta > 1$. An analytic understanding of the relation (\ref{Eq:Jn})   
is so far only available for the case without selection, where 
$1/q$ increase linearly with population size $N$ \cite{W1982A,W1982B}. 
We note, finally, that the time series of fixation events has interesting 
statistical properties \cite{G2001,PK2007}, which are however outside the scope
of the present article.

\section{\label{Sec:Out} Summary and outlook}

In this paper we have reviewed some aspects of evolutionary dynamics in the arguably simplest 
setting: A population of fixed size $N$ evolving in a time-independent environment, supplied by 
independently acting beneficial mutations at a constant rate $U$. The quantity of primary interest
is the speed $v_N$ of logarithmic fitness increase, which is determined by the parameters
$N$ and $U$ and by the probability distribution $g(s)$ of mutational effects with typical
scale $s_b$. 

On a qualitative level, one finds three distinct evolutionary regimes. For small populations, 
in the sense of (\ref{onset}), beneficial mutations are well separated in time and 
sweep through the population independently. As a consequence, the speed $v_N$ is proportional
to $N U$. For larger populations the clones generated by different mutations interfere and the
increase of the speed is only logarithmic in $N$. Finally, in the limit of infinite populations,
the speed saturates to a finite value (for the discrete time WF model). 
In the last regime the problem can be solved exactly, but, due
to a conspiracy of the small parameters $U$ and $s_b$, this description applies only
to hyperastronomically large populations (see Eq.~(\ref{Nc})).  Real microbial populations
of the kind used in evolution experiments typically operate in the intermediate clonal interference
regime, which has been the main focus of the article. 

Most work on the finite 
population problem has considered the case of a single selection coefficient (model I), 
where fitness is discrete. This offers considerable advantages for both approximate and rigorous
analytic studies as well as for numerical simulations, which are able to exlore the asymptotic
regime where $\ln N$ (and not just $N$) is large. A summary of the present state of affairs with
regard to analytic approximations for the speed is given in Fig.~\ref{Fig:comall}. The case of a 
continuous distribution of selection coefficients (model II) is less well understood. Despite its conceptual shortcomings, the Gerrish-Lenski approximation provides a quantitatively
rather satisfactory description of the speed over the experimentally relevant range of population
sizes (see Fig.~\ref{Fig:GL}), although it fails completely when the infinite population limit
is approached. We have argued above that model I should provide a lower bound on the speed of
evolution for general distributions of selection coefficients, which implies that the speed increases at least as fast as $\ln N$ also for model II, and possibly faster for distributions $g(s)$ that decay
more slowly than exponentially.

The unifying paradigm used throughout the article is the description of the evolutionary process
in terms of a traveling wave of constant shape moving along the fitness axis 
\cite{Tsimring1996,Rouzine2003}. This idea has proved to be successful also in the related
but distinct context of competitive evolution, 
where selection is decoupled from reproduction \cite{Peng2003,Kloster2005}. 
Competitive evolution models mimic a process of artificial (rather than natural) selection, where
the character (``trait'') of the types that is being selected is not the reproductive ability (fitness) 
of individuals. In one variant, individuals are assigned a scalar trait
which is handed on to the offspring subject to random mutations. In one round of reproduction,
each individual creates the same number of offspring,
and subsequently the $N$ with the largest value of the trait are selected for the next round
\cite{Brunet2007}. This model falls into the large class of noisy traveling waves of 
Fisher-Kolmogorov type \cite{Saarloos2003,Panja2004,Brunet2006}, which are much better understood
than the problems described in the present article. Apart from accurate analytic approximations
to the wave speed, also the genealogies of populations can be 
addressed, which display an interesting relation to the
statistical physics of disordred systems \cite{Brunet2006}. 
In contrast, the genealogical properties of the WF model with
selection are largely unknown. 

Although the models described here are of considerable interest for the interpretation of 
evolutionary experiments \cite{Desai2007,EL2003,Hegreness2006,Perfeito2007,RVG2002,Tsimring1996,dVR2006}, 
the reader should not be left with the impression that
they provide a description that is realistic in all or even most respects. For example,
the assumption of a constant supply of beneficial mutations cannot be true at arbitrarily
long times, and indeed the rate of fitness increase is generally observed to slow down in experiments
\cite{dVZGBL1999}. One way to take this effect into account is by modeling the genotype as a sequence
with a finite number of sites at which mutations can take place \cite{KO2005}. 

Another approach,
known as Kingman's house-of-cards model \cite{K1978},
retains the infinite number of sites approximation but modifies the basic mutation step
(\ref{Eq:multiplicative}) 
such that the mutant fitness $w_i'$ itself is drawn from a fixed fitness distribution
$\tilde g(w)$. The probability of chosing a beneficial mutation with $w_i' > w_i$ then
decreases as the mean fitness grows, and correspondingly the logarithmic fitness increases
in a sublinear manner determined by the tail of $\tilde g$ \cite{Park2008}. 
In fact this problem turns out
to be simpler than the one discussed in the present article, because the diminishing rate of 
beneficial mutations $(U \to 0$) drives the system into the periodic selection regime
where selective sweeps can be treated as independent. 

Kingman's assumption that the fitness of 
the offspring is uncorrelated with that of the parent is hardly more realistic than
the assumption of independent fitness effects of different mutations which underlies
Eq.~(\ref{Eq:multiplicative}). The few examples available so far indicate that real
fitness landscapes lie between these two extremes \cite{PKWT2007,deVisser2009}, which 
implies that the structure of the type space cannot be ignored. Like the modeling of 
evolutionary dynamics which we have discussed in this article, the mathematical characterization
of such fitness landscapes offers a host of challenging problems that can be fruitfully explored by 
statistical physicists.

\begin{acknowledgements}

This work was supported by DFG within SFB 680, and by the Alexander von Humboldt foundation through a fellowship
to DS. We acknowledge the kind hospitality of the Lorentz Center, Leiden, where part of this paper was written.

\end{acknowledgements}



\appendix

\section{Simulating the Wright-Fisher model}
This Appendix is devoted to explaining how we simulated model I for
population sizes up to $10^{300}$, as displayed in
Fig.~\ref{Fig:comall}. The algorithm is based on that of
\cite{PK2007}, which we describe first.  
As in Sec.~\ref{Sec:Inf_Cal}, we denote 
the frequency of individuals with fitness $\e^{n s_b}$ at generation
$t$ by $f_t(n)$.
Assume that at time $t$ there are $k+1$ distinct fitness values
present in the population, i.e. $f_t(n) = 0$ if $n \le n_0$ or $n > n_0 + k$.
It is straightforward to see from Eq.~\eqref{eq:wright-fisher-process} that 
the number $m_i$ of individuals having $n_i \equiv n_0+i$ ($i=1,\ldots,k+1$)
mutations at generation $t+1$ is determined by the multinomial distribution 
\begin{equation}
p(m_1,\ldots,m_{k+1}) = N! \prod_{i=1}^{k+1} \frac{p_i^{m_i}}{m_i!},
\label{Eq:multi}
\end{equation}
where 
\begin{equation}
p_i = f_t(n_i) (1-U) \frac{\e^{n_i s_b}}{\bar w(t)}
+ f_t(n_i-1) U \frac{\e^{(n_i -1)s_b}}{\bar w(t)}.
\end{equation}
Note that the effect of mutations is already implemented in
the above algorithm, 
which is equivalent to the WF model in Section~\ref{Sec:model} 
(first selection then mutation).
Since this multinomial distribution can be written as
\begin{equation}
p(m_1,\ldots,m_{k+1}) =\prod_{j=2}^{k+1} \binom{N_i}{m_i} (1-q_i)^{N_i-m_i} q_i^{m_i},
\end{equation}
where 
\begin{equation}
q_i = \frac{p_i}{\sum_{j=1}^i p_j},
\end{equation}
$N_i = N_{i+1}-m_{i+1}$ and $N_{k+1} = N$, the multinomially
distributed random numbers can be generated by drawing 
binomial random numbers $k$ times.
To be specific, we first draw $m_{k+1}$ from the distribution
\begin{equation}
\binom{N}{m_{k+1}}(1-q_{k+1})^{N-m_{k+1}} q_{k+1}^{m_{k+1}},
\end{equation}
then the $m_{j}$ are determined in the order of $j=k$, $k-1$, $\ldots$, 2 
by the conditional distribution
\begin{equation}
\binom{N_j}{m_{j}}(1-q_{j})^{N_j-m_{j}} q_{j}^{m_{j}}.
\end{equation}
Finally, $m_1$ is given by $N_1 = N - \sum_{j=2}^{k+1} m_j$.

Since it is not possible to generate integers as large as $10^{100}$
in present day computers, in our simulations of very large populations
we treat the $m_j$ as real numbers.
To be specific, we use the following algorithm.
If $N_j < 10^9$, we generate binomially distributed integer random
variables. If $N_j > 10^9$,
we first check if the mean $\bar m_j \equiv N_j q_j$ is larger than 
prescribed number $M$ which was set as 100 in our
simulations\footnote{The results do not depend on this choice.}.
If $\bar m_j < M$, we generate Poisson distributed random numbers with
mean $\bar m_j$. Since $N_j$ is sufficiently large and $q_j < 10^{-7}$, the Poisson
distribution accurately approximates the binomial distribution 
in this situation. On the other hand, if $\bar m_j > M$, we invoked
the central limit theorem to approximate
the binomial distribution by a Gaussian; that is,
$m_j = \bar m_j + \sqrt{N_j q_j (1-q_j)} N(0,1)$, where
$N(0,1)$ is a normally distributed random number with mean 0 and variance 1. 

Needless to say, the above algorithm is successful up to
hyperastronomical population sizes because
the fitness space is quantized and the number of possible fitness values
at each generation, determined by the lead $L_0$, increases only 
as $\sim \ln N$. The direct
application of this method to model II is not feasible, 
because in that case the number of different fitness values is at least $N U$.

\end{document}